\documentclass[english,reprint,aps,prb,superscriptaddress,showpacs,showkeys,twocolumn]{revtex4-1}
\usepackage[T1]{fontenc}
\usepackage[utf8]{inputenc}
\usepackage{xcolor}
\usepackage{pdfcolmk}
\usepackage{babel}
\usepackage{array}
\usepackage{booktabs}
\usepackage{units}
\usepackage{textcomp}
\usepackage{amstext}
\usepackage{graphicx}
\PassOptionsToPackage{normalem}{ulem}
\usepackage{ulem}
\usepackage[unicode=true,
 bookmarks=true,bookmarksnumbered=false,bookmarksopen=false,
 breaklinks=false,pdfborder={0 0 1},backref=false,colorlinks=true]
 {hyperref}
\hypersetup{pdftitle={Magnetoelastic phenomena in antiferromagnetic uranium intermetallics:the UAu2Si2 case},
 pdfauthor={M.Valiska,  H. Saito, T. Yanagisawa, Ch. Tabata, H. Amitsuka, K. Uhlirova, J. Prokleska, P. Proschek, J. Valenta, M. Misek, D.I. Gorbunov, J. Wosnitza, and V. Sechovsky},
 pdfkeywords={thermal expansion, magnetostriction, magnetization, tricritical point}}

\makeatletter

\providecommand{\tabularnewline}{\\}
\providecolor{lyxadded}{rgb}{0,0,1}
\providecolor{lyxdeleted}{rgb}{1,0,0}

\DeclareRobustCommand{\lyxsout}[1]{\ifx\\#1\else\sout{#1}\fi}

\makeatother

\begin{document}

\title{Magnetoelastic phenomena in antiferromagnetic uranium intermetallics:\\
 the $\mathrm{UAu_{2}Si_{2}}$ case}

\author{M.Vali\v{s}ka}
\email{michal.valiska@gmail.com}

\affiliation{Faculty of Mathematics and Physics, Charles University, DCMP, Ke
Karlovu 5, CZ-12116 Praha 2, Czech Republic}

\affiliation{Institut Laue Langevin, 71 Avenue des Martyrs, CS 20156, F-38042
Grenoble Cedex 9, France}

\author{H. Saito}

\affiliation{Graduate School of Science, Hokkaido University, Sapporo 060-0810,
Japan}

\author{T. Yanagisawa}

\affiliation{Graduate School of Science, Hokkaido University, Sapporo 060-0810,
Japan}

\author{Ch. Tabata}

\affiliation{Institute of Materials Structure Science, High Energy Accelerator
Organization, Tsukuba 305-0801, Japan }

\author{H. Amitsuka}

\affiliation{Graduate School of Science, Hokkaido University, Sapporo 060-0810,
Japan}

\author{K. Uhl\'{i}\v{r}ov\'{a} }

\affiliation{Faculty of Mathematics and Physics, Charles University, DCMP, Ke
Karlovu 5, CZ-12116 Praha 2, Czech Republic}

\author{J. Prokle\v{s}ka}

\affiliation{Faculty of Mathematics and Physics, Charles University, DCMP, Ke
Karlovu 5, CZ-12116 Praha 2, Czech Republic}

\author{P. Proschek }

\affiliation{Faculty of Mathematics and Physics, Charles University, DCMP, Ke
Karlovu 5, CZ-12116 Praha 2, Czech Republic}

\author{J. Valenta}

\affiliation{Faculty of Mathematics and Physics, Charles University, DCMP, Ke
Karlovu 5, CZ-12116 Praha 2, Czech Republic}

\author{M. M\'{i}\v{s}ek}

\affiliation{Institute of Physics, ASCR v.v.i, Na Slovance 2, 182 21, Prague,
Czech Republic}

\author{D.I. Gorbunov}

\affiliation{Hochfeld-Magnetlabor Dresden (HLD-EMFL), Helmholtz-Zentrum Dresden-Rossendorf,
01328 Dresden, Germany}

\author{J. Wosnitza}

\affiliation{Hochfeld-Magnetlabor Dresden (HLD-EMFL), Helmholtz-Zentrum Dresden-Rossendorf,
01328 Dresden, Germany}

\affiliation{Institut für Festkörperphysik und Materialphysik, TU Dresden, 01062
Dresden, Germany }

\author{V. Sechovský}

\affiliation{Faculty of Mathematics and Physics, Charles University, DCMP, Ke
Karlovu 5, CZ-12116 Praha 2, Czech Republic}
\begin{abstract}
Thermal expansion, magnetostriction and magnetization measurements
under magnetic field and hydrostatic pressure were performed on a
$\mathrm{UAu_{2}Si_{2}}$ single crystal. They revealed a large anisotropy
of magnetoelastic properties manifested by prominent length changes
leading to a collapse of the unit-cell volume accompanied by breaking
the fourfold symmetry (similar to that in $\mathrm{URu_{2}Si_{2}}$
in the hidden-order state) in the antiferromagnetic state as consequences
of strong magnetoelastic coupling. The magnetostriction curves measured
at higher temperatures confirm a bulk character of the $\unit[50]{K}$
weak ferromagnetic phase. The large positive pressure change of the
ordering temperature predicted from Ehrenfest relation contradicts
the more than an order of magnitude smaller pressure dependence observed
by the magnetization and specific heat measured under hydrostatic
pressure. A comprehensive magnetic phase diagram of $\mathrm{UAu_{2}Si_{2}}$
in magnetic field applied along the $c$ axis is presented. The ground-state
antiferromagnetic phase is suppressed by a field-induced metamagnetic
transition that changes its character from the second to the first
order at the tricritical point. 
\end{abstract}


\maketitle

\section{Introduction}

The anharmonic lattice vibrations due to the asymmetric bonding potential
lead to increasing equilibrium interatomic distances with rising temperature
in solids. The corresponding thermal expansion is a monotonously increasing
function of temperature. The anisotropy of bonding within the crystal
lattice causes the anisotropy of thermal expansion which is manifested
by different temperature dependences of the linear thermal expansion
$\left(\Delta l/l\right)_{i}$ along the different crystallographic
axes, $i$. The thermal expansion of metals includes also a conduction-electron
contribution. This plays a considerable role at low temperatures where
the phonon term almost vanishes. 

The magneto-structural coupling, reflecting the interplay between
the spin and lattice degrees of freedom, brings additional contributions
to the thermal expansion in magnetic compounds. The magnetocrystalline
anisotropy leads to anisotropic magnetic contributions to the thermal
expansion. Magnetic materials then exhibit unusual thermal-expansion
behavior especially in a magnetically ordered state. 

The thermal expansion, similar to the specific heat, is thus a useful
probe for investigations of thermodynamic phenomena in magnetic materials
(without applying magnetic field). The specific heat has only bulk
character, whereas the thermal expansion enables us to study also
the anisotropy of thermodynamic properties.

The spatially extended uranium 5$f$-electron wave functions in solids
considerably interact with the overlapping 5$f$ orbitals of the nearest-neighbor
U ions and the 5$f$-electron states hybridize with valence-electron
states of non-uranium ligands (5$f$-ligand hybridization\cite{Koelling1985})
and the 5$f$ electrons even participate in bonding.\cite{Smith1983,Eriksson1991}
The exchange interactions that are coupling the uranium 5$f$-electron
magnetic moments in U antiferromagnets are strongly anisotropic. The
direct exchange interactions are due to the 5$f$-5$f$ orbitals overlap.
The anisotropy of these, typically ferromagnetic (FM), interactions
and as well as the magnetocrystalline anisotropy are determined by
the arrangement of the nearest-neighbor U ions in the lattice. The
antiferromagnetic (AFM) interactions in U compounds are usually mediated
by the anisotropic 5$f$-ligand hybridization. The magnetoelastic
coupling then produces highly anisotropic magnetic contributions to
the thermal expansion and magnetostriction, especially in U antiferromagnets.

The anisotropy of magnetoelastic phenomena is a subject of numerous
papers on U magnetics. Most frequently they have been dedicated to
the intriguing properties of $\mathrm{URu_{2}Si_{2}}$, the most thoroughly
studied uranium compound in more than the last three decades. There
were no doubts about bulk superconductivity in $\mathrm{URu_{2}Si_{2}}$
below $\unit[1.5]{K}$ since the earliest stage of research of this
compound. Interpretation of the huge specific-heat peak and the Cr-like
anomaly of electrical resistivity both at $\unit[17.5]{K}$ were,
however, always a subject of dispute. First, these were interpreted
in terms of a transition to a weak itinerant antiferromagnetism,\cite{Palstra1985}
static charge-density wave (CDW) or spin-density wave (SDW) transition\cite{Maple1986}
or a local U-moment antiferromagnetism.\cite{Schlabitz1986} No long-range
magnetic order nor any sign of a static CDW or SDW formation below
$\unit[17.5]{K}$ has been confirmed by microscopic methods, however.
Within time the term ‘‘hidden order’’ (HO) was introduced to describe
the unknown ordered state which allows unconventional superconductivity
to occur at $\unit[1.5]{K}$. $\mathrm{URu_{2}Si_{2}}$ exhibits a
non-magnetic, non-structural HO phase transition at $T_{\mathrm{HO}}=\unit[17.5]{K}$
where the order parameter and elementary excitations so far could
not be determined by microscopic experiments and only dynamical magnetic
fluctuations are observed. Comprehensive information on physics of
$\mathrm{URu_{2}Si_{2}}$ can be found in Ref.\cite{Mydosh2011} and
references therein. The volume of this compound reduces considerably
below $T_{\mathrm{HO}}$ as manifested by a sharp positive peak in
the thermal-expansion coefficient at $T_{\mathrm{HO}}$. The volume
reduction of the tetragonal structure is due to the basal-plane shrinkage.
The simultaneous lattice expansion along the $c$ axis is too small
to compensate the negative basal-plane effect.\cite{deVisser1986,deVisser1990,Motoyama2008,Ran2016}
The possibility of a slight orthorhombic distortion of the tetragonal
lattice at temperatures below $T_{\mathrm{HO}}$ plays an important
role in the physics of $\mathrm{URu_{2}Si_{2}}$. We will come back
to this point in the Discussion section. 

The influence of anisotropic exchange interactions on the anisotropy
of thermal expansion in U antiferromagnets is manifested by the magnetoelastic
behavior of two other $\mathrm{U}T_{2}X_{2}$ ($T$ – transition metal,
$X$ – $p$-electron metal) compounds with the tetragonal $\mathrm{ThCr_{2}Si_{2}}$
structure, $\mathrm{UCo_{2}Si_{2}}$ \cite{Andreev2017,Andreev2013}
and $\mathrm{UNi_{2}Si_{2}}$ \cite{Honda1999} and several antiferromagnets
from the family of hexagonal $\mathrm{U}TX$ compounds crystalizing
in the $\mathrm{ZrNiAl}$ structure. In the both structures the nearest
U-U neighbors are located in the basal plane where the U magnetic
moments are coupled ferromagnetically. All these compounds exhibit
the strong uniaxial anisotropy fixing the U moments to the $c$ axis,
which is the easy magnetization direction. The AFM structures in these
materials are built of the FM basal-plane layers antiferromagnetically
coupled along the $c$ axis. The thermal expansion below the Néel
temperature ($T_{\mathrm{N}}$) in these antiferromagnets (similar
to $\mathrm{URu_{2}Si_{2}}$ below $T_{\mathrm{HO}}$) is strongly
anisotropic as well as the magnetostriction accompanying field-induced
metamagnetic transitions from the AFM to paramagnetic state. The corresponding
$a$- and $c$-axis linear thermal expansions $\left(\Delta l/l\right)_{a}$
and $\left(\Delta l/l\right)_{c}$, respectively have in all cases
opposite signs. The volume thermal expansion calculated according
to 

\begin{equation}
\Delta V/V=2\cdot\left(\Delta l/l\right)_{a}+\left(\Delta l/l\right)_{c}\label{eq:volume}
\end{equation}
 for the $\mathrm{U}T_{2}X_{2}$ compounds is small as a result of
compensation of the opposite-sign linear expansions. The linear magnetostrictions
$\lambda_{a}$ and $\lambda_{c}$ accompanying a metamagnetic transition
are also of opposite signs leading to small volume magnetostriction.
However, they have opposite polarities with respect to the corresponding-direction
of thermal expansions. In fact, the magnetic contributions to thermal
expansion of an antiferromagnet below $T_{\mathrm{N}}$ are suppressed
by the opposite polarity corresponding to magnetostrictions caused
by the metamagnetic transition. 

$\mathrm{UAu_{2}Si_{2}}$ belongs to the family of $\mathrm{U}T_{2}\mathrm{Si_{2}}$
compounds which adopt the tetragonal $\mathrm{ThCr_{2}Si_{2}}$ structure
($\mathrm{UIr_{2}Si_{2}}$ and $\mathrm{UPt_{2}Si_{2}}$ crystallize
in the $\mathrm{CaBe_{2}Ge_{2}}$ structure\cite{Sechovsky1998}).
These compounds exhibit a spectrum of physical properties ranging
from Pauli paramagnets ($\mathrm{UFe_{2}Si_{2}}$,\cite{Szytula1988}
$\mathrm{URe_{2}Si_{2}}$, and $\mathrm{UOs_{2}Si_{2}}$\cite{Palstra1986})
to magnetically ordered systems which are mostly complex and either
AFM ($\mathrm{UCr_{2}Si_{2}}$,\cite{Matsuda2003} $\mathrm{UCo_{2}Si_{2}}$,\cite{Chelmicki1985}
$\mathrm{UNi_{2}Si_{2}}$,\cite{Chelmicki1983} $\mathrm{URh_{2}Si_{2}}$,\cite{Ptasiewicz1981}
$\mathrm{UPd_{2}Si_{2}}$,\cite{Ptasiewicz1981,Honma1998} $\mathrm{UIr_{2}Si_{2}}$,\cite{Palstra1986,Dirkmaat1990}
and $\mathrm{UPt_{2}Si_{2}}$ \cite{Palstra1986}), or FM ($\mathrm{UMn_{2}Si_{2}}$\cite{Szytula1988}).
$\mathrm{UCu_{2}Si_{2}}$ \cite{Chelmicki1985,Matsuda2005,Honda2006,Troc2012}
exhibits a FM ground state with an additional AFM phase at higher
temperatures. An exceptional case among them is the well-known $\mathrm{URu_{2}Si_{2}}$\cite{Palstra1985}
exhibiting the hidden-order transition. The magnetism of $\mathrm{UAu_{2}Si_{2}}$
was for many years left unclear mainly due to metallurgical difficulties.\cite{Sechovsky1998}

Quite recently we have successfully prepared a $\mathrm{UAu_{2}Si_{2}}$
single crystal and commenced systematic investigations of its intrinsic
properties. The results obtained by measurements of magnetization,
specific heat and electrical resistivity\cite{Tabata2016} followed
by neutron diffraction \cite{Tabata2017} and $^{29}$Si-NMR\cite{Tabata2017a}
experiments corroborate the conclusion about the ground state of $\mathrm{UAu_{2}Si_{2}}$
as an uncompensated antiferromagnet, contrary to previous reports
on polycrystals.\cite{Palstra1986,Rebelsky1991,Torikachvili1992,Lin1997,Saran1998}
$\mathrm{UAu_{2}Si_{2}}$ undergoes a FM-like transition at $\unit[50]{K}$
(referred to as $T_{2}$) followed by another magnetic phase transition
($T_{\mathrm{m}}$) around $\unit[20]{K}$. All our previous measurements
show a large magnetocrystalline anisotropy with the direction of magnetic
moments along the $c$ axis. The propagation vector is $\left(\nicefrac{2}{3},0,0\right)$
and the magnetic structure can be described as a stacking sequence
(+ + -) of the FM $ac$ plane sheets along the $a$ axis.\cite{Tabata2017}
The specific-heat measurements point to an enhanced value of the Sommerfeld
coefficient $\gamma\sim\unit[180]{mJ\,K^{-2}\,mol^{-1}}$. Now, we
investigated the $\mathrm{UAu_{2}Si_{2}}$ single crystal by use of
thermal-expansion, magnetostriction and magnetization measurements
up to high magnetic fields and under hydrostatic pressure. The results
of the present paper, are complementary to our previous neutron-diffraction
work, confirming the uncompenasted antiferromagnetic (UAFM) ground
state of the compound,\cite{Tabata2017} and bring evidence for the
intrinsic nature of the FM component below $\sim\unit[50]{K}$ previously
reported as parasitic in our first single-crystal study.\cite{Tabata2016}
Magnetization measurements in pulsed high magnetic fields helped us
to complete the phase diagram of $\mathrm{UAu_{2}Si_{2}}$ and revealed
signs of the presence of a tricritical point (TCP). 

\section{Experimental Details}

The $\mathrm{UAu_{2}Si_{2}}$ single crystal used in this study was
prepared using the floating-zone method in an optical furnace (Crystal
Systems Co.) in a similar way as in our previous work.\cite{Tabata2016}
Nevertheless, to obtain a higher-quality and larger single crystals
we have optimized the entire growth process. The initial polycrystalline
rod with diameter of $\unit[6]{mm}$ and length of $\sim\unit[100]{mm}$
was prepared from the starting elements of U (initially $99.9\%$
and consequently purified by the Solid State Electrotransport Method
under ultra high vacuum\cite{Haga1998}), Au ($99.99\%$) and Si ($99.999\%$).
The rod was subsequently annealed at $1000\text{\textdegree C}$ for
three days, cut in two parts and placed in the optical furnace, where
the shorter bottom part served as a polycrystalline seed and the main
larger rod hung from the top as feed material for the growth. The
chamber of the optical furnace was evacuated to $\sim\unit[10^{-6}]{mbar}$
and the growth itself was done under the protective Ar atmosphere
with a flow of $\unit[1.5]{l\,min^{-1}}$ in an overpressure of $\sim\unit[0.2]{MPa}$.
The power of the lamps in the furnace was adjusted to keep the temperature
of the hot zone slightly above the melting point. Both, the seed and
feed rod were slowly pulled through the hot zone with a speed of $\unit[1]{mm\,h^{-1}}$
and without rotation. The quality of the grown single crystal was
checked by the x-ray Laue method and Energy-dispersive x-ray (EDX)
analysis. 

Length changes were measured using a miniature capacitance dilatometer\cite{Rotter2007}
mounted in PPMS 9 T and PPMS 14 T (Quantum Design Co.) between $\unit[2]{K}$
and $\unit[100]{K}$ in magnetic fields up to $\unit[14]{T}$. Magnetization
measurements in static fields were done using the VSM option (Vibrating
Sample Magnetometer) implemented in the PPMS 14 T. Specific heat measurements
up to $\unit[9]{T}$ were performed using the relaxation technique
by PPMS 9 T (Quantum Design Co.).

The magnetization in pulsed magnetic fields up to $\sim\unit[58]{T}$
was measured at the Dresden High Magnetic Field Laboratory using a
coaxial pick-up coil system. The high-field magnetometer is described
in Ref. \cite{Skourski2011}. Absolute values of the magnetization
were calibrated using static-field measurements. 

The magnetization measurements under hydrostatic pressure were performed
in the MPMS SQUID (Quantum Design Co.) magnetometer using a CuBe pressure
cell\cite{Kamarad2004} with a liquid pressure medium and a piece
of lead as manometer. The heat capacity of the $\mathrm{UAu_{2}Si_{2}}$
sample under high pressures was measured by the means of steady-state
calorimetry.\cite{Sullivan1968} Double-layered CuBe/NiCrAl piston-cylinder
pressure cell was used to generate pressures up to $\sim\unit[3]{GPa}$,
with a Daphne 7373 pressure medium and a manganin manometer. A micro
strain-gauge was used for periodic heating of the sample and an Au/AuFe
thermoucouple was used to measure it’s temperature oscillations. The
amplitude of oscillations is inversely proportional to the sample
heat capacity. Technical details of the method\cite{Sullivan1968,Kraftmakher2004}
and actual experimental setup used in our experiments \cite{Misek2013}
are beyond the scope of this paper and can be found elsewhere.

\section{Results}

\subsection{Thermal-expansion measurements }

\begin{figure}
\begin{centering}
\includegraphics[scale=0.33]{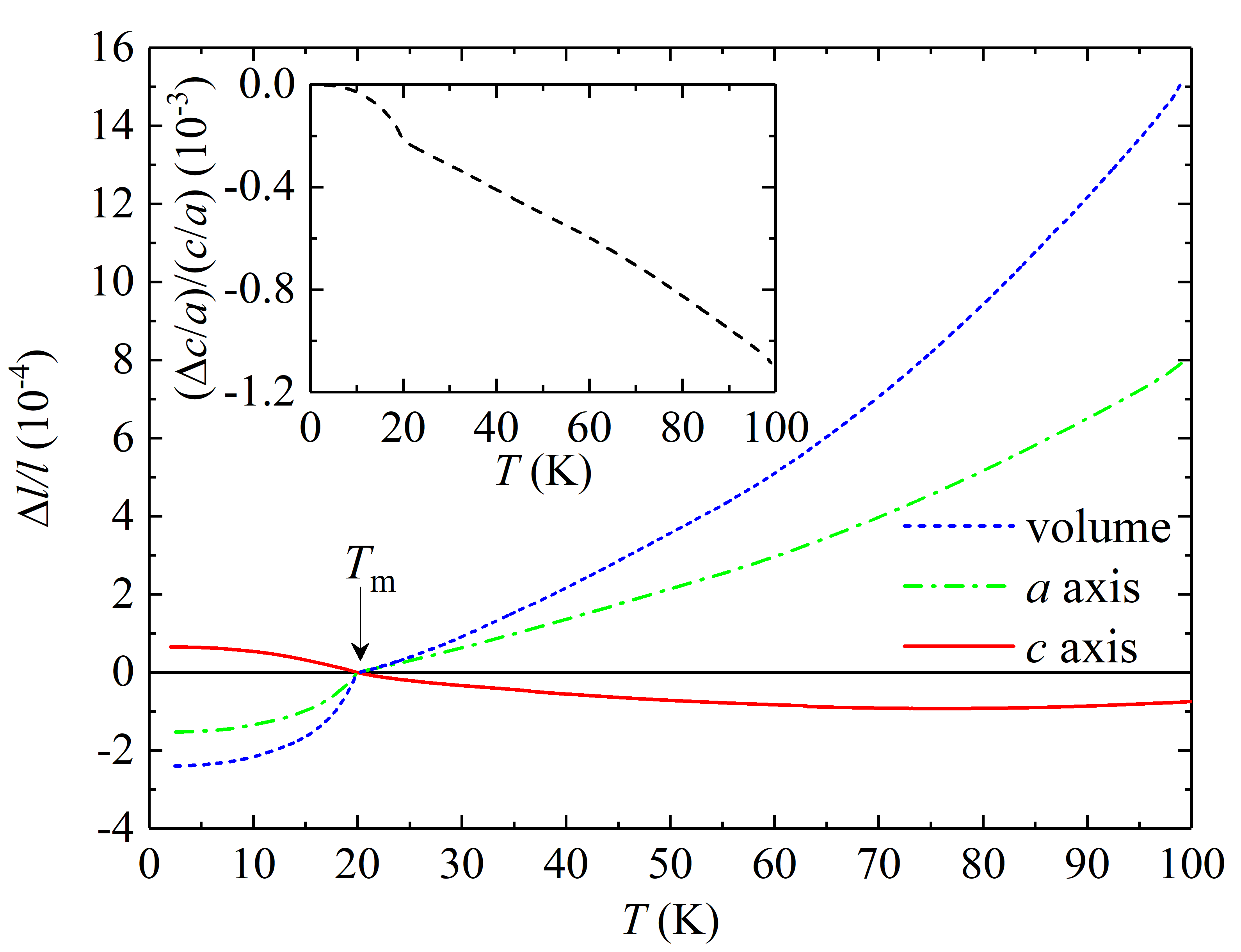}
\par\end{centering}
\caption{\label{fig:The-linear-thermal}Linear thermal expansion for the $a$
and $c$ axis and volume change measured without applied external
magnetic field. The inset shows $c/a$ as a function of temperature.
The curve is normalized to be equal to $0$ at $\unit[2]{K}$.}
\end{figure}

The $a$- and $c$-axis linear thermal expansions together with the
volume expansion calculated according to Eq. (\ref{eq:volume}) are
plotted in Fig. \ref{fig:The-linear-thermal}. The individual data
are vertically shifted to set them equal to $0$ at $T_{\mathrm{m}}=\unit[20.5]{K}$.
Below this magnetic phase transition we observe a significant change
in the temperature dependence of the thermal expansion in line with
our previous reports.\cite{Tabata2016,Tabata2017}

There obviously is a large anisotropy of the thermal expansion over
the entire temperature range. The linear thermal expansion along the
$a$ axis shows a continuous decrease from $\unit[100]{K}$ down to
the ordering temperature $T_{\mathrm{m}}=\unit[20.5]{K}$, where it
bends down rapidly pointing at a large contraction of the unit cell
along the $a$ axis ($-1.5\times10^{-4}$ between $T_{\mathrm{m}}$
and $\unit[2]{K}$). On the other hand, the thermal expansion along
the $c$ axis shows a broad minimum around $\unit[75]{K}$ followed
by an increase at lower temperatures. The ordering temperature appears
as an inflection point and the $c$-axis expansion below $\unit[20.5]{K}$
is $6.6\times10^{-5}$. The volume thermal expansion is calculated
using Eq. (\ref{eq:volume}). The obtained relative volume change
shows a large reduction below $T_{\mathrm{m}}$($-2.3\times10^{-4}$
between $T_{\mathrm{m}}$ and $\unit[2]{K}$) as shown in Fig. \ref{fig:The-linear-thermal}.
The continuous character of the linear thermal expansion at $T_{\mathrm{m}}$
points to a second-order transition. 

\begin{figure}
\begin{centering}
\includegraphics[scale=0.33]{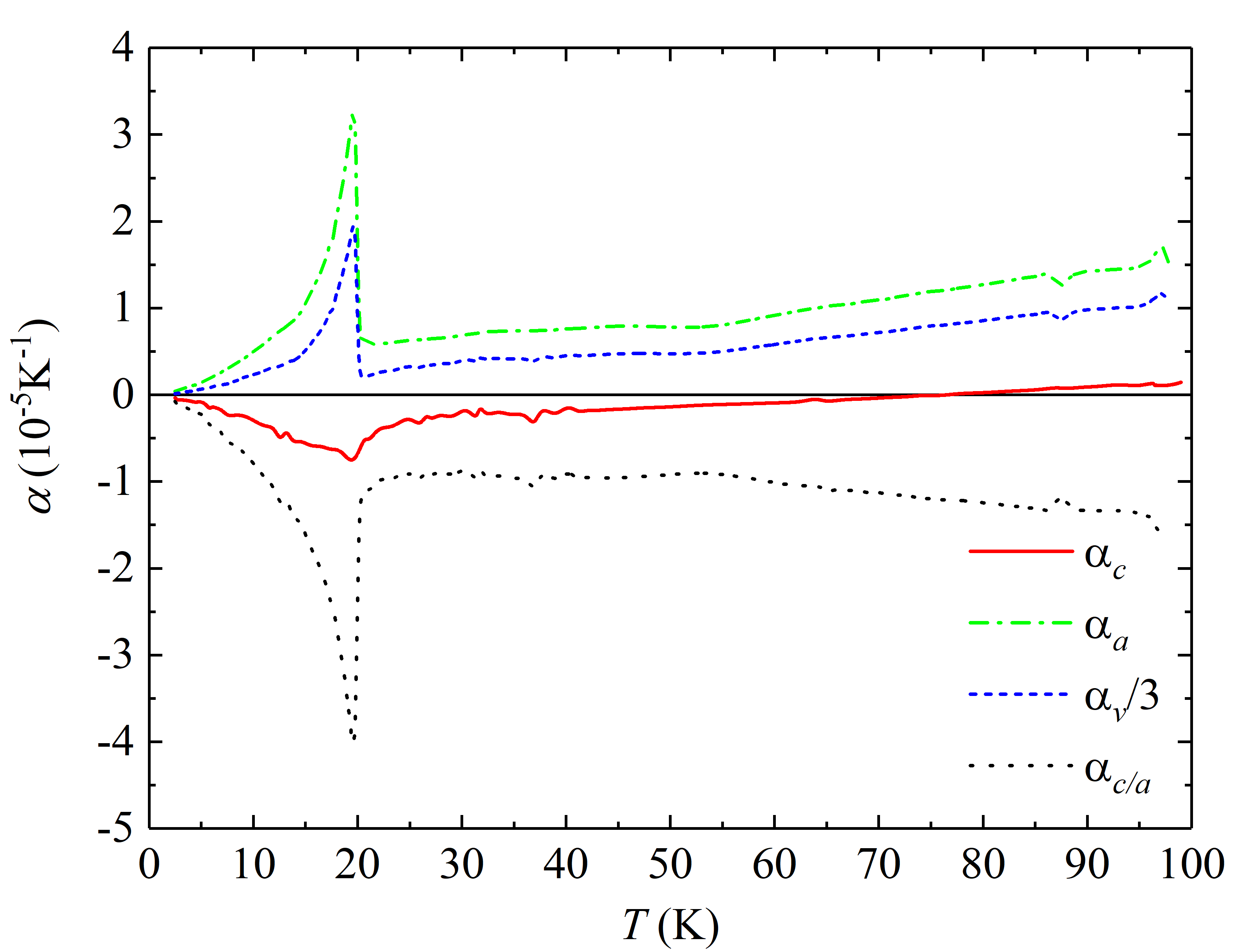}
\par\end{centering}
\caption{\label{fig:The-linear-thermal-1}Linear thermal-expansion coefficients
for the $a$ and $c$ axis together with the $\alpha_{\mathrm{c/a}}$.
Note that the volume dependence is plotted as $\alpha_{v}/3$. }

\end{figure}

The linear thermal-expansion coefficients $\alpha_{i}$ are defined
as temperature derivatives of the linear thermal expansion $\left(\Delta l/l\right)_{i}$,
i.e., $\alpha_{i}=\mathrm{d}\left(\Delta l/l\right)_{i}/\mathrm{d}T$.
The calculated linear thermal-expansion coefficients are plotted in
Fig. \ref{fig:The-linear-thermal-1} together with the volume thermal-expansion
coefficient defined as $\alpha_{v}=2\alpha_{a}+\alpha_{c}$. It is
also useful to determine the temperature dependence of the $c/a$
ratio. We, therefore, define the following temperature coefficient
$\alpha_{c/a}=\alpha_{c}-\alpha_{a}$. Both quantities are plotted
in Fig. \ref{fig:The-linear-thermal-1}.

The linear thermal-expansion coefficient for the $c$ axis becomes
negative below $\unit[75]{K}$ where the minimum of its relative length
change occurs. There are sharp peaks at $T_{\mathrm{m}}$ for all
measured curves and a small change of the slope above $\unit[50]{K}$
visible in $\alpha_{a}$ coefficient that is projected in the remaining
computed quantities. Integration of the $\alpha_{c/a}$ coefficient
along the whole temperature range results in the relative temperature
dependence of the $c/a$ ratio, see inset in Fig. \ref{fig:The-linear-thermal}. 

The temperature dependence of this ratio is monotonous and decreasing
nearly linear above $T_{\mathrm{m}}$. The slope increases below the
$T_{\mathrm{m}}$, emphasizing again the prominent contraction of
the $a$ axis.

As will be discussed below, ultra-pure samples of the isostructural
compound $\mathrm{URu_{2}Si_{2}}$ studied by synchrotron x-ray diffraction
show a small orthorhombic distortion when entering the hidden-order
state.\cite{Tonegawa2014} The size of the distortion and/or sample
quality are possible reasons why it was not observed by the thermal-expansion
measurements. \cite{deVisser1986,Kuwahara1997}

\begin{figure}
\begin{centering}
\includegraphics[scale=0.33]{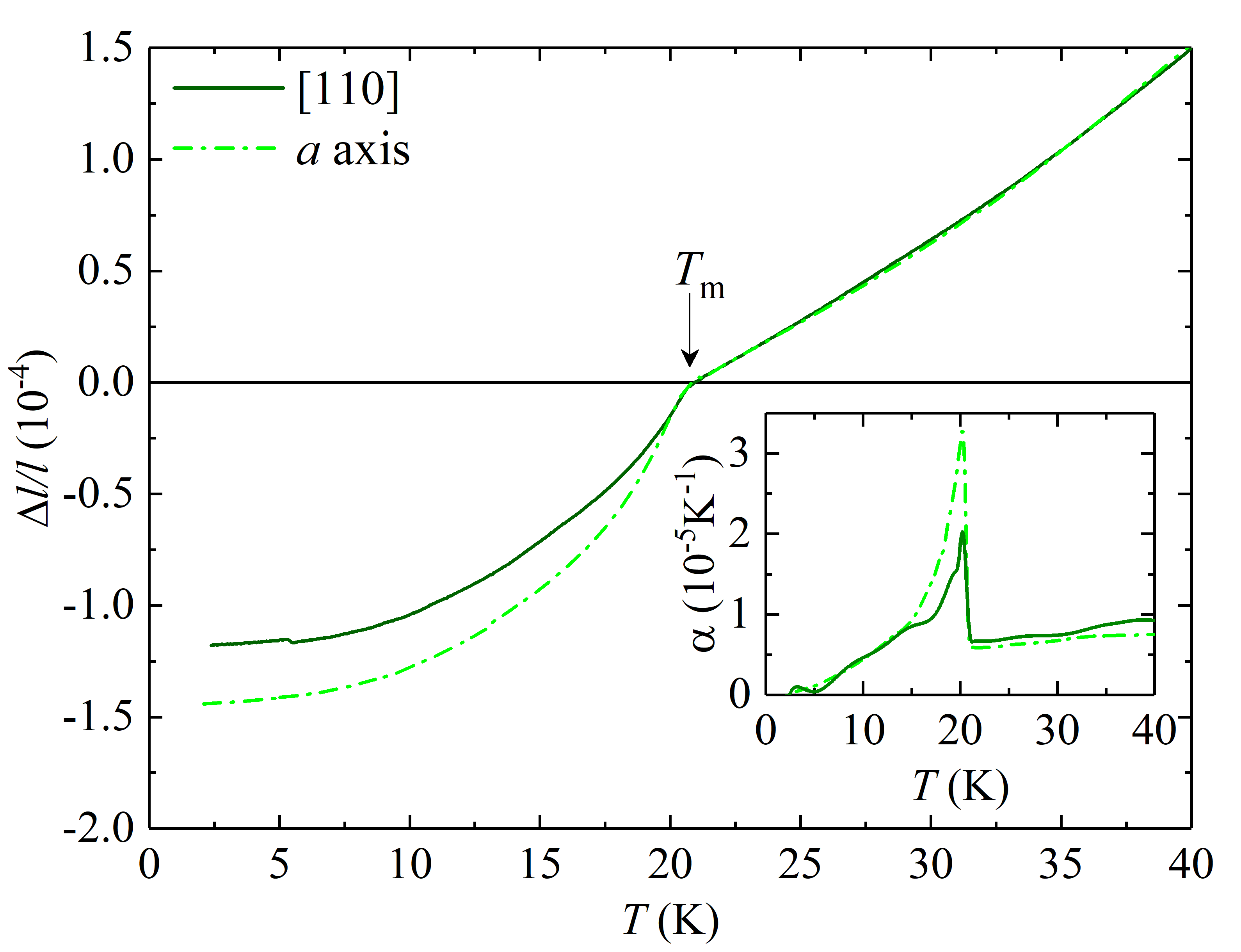}
\par\end{centering}
\caption{\label{fig:The-linear-thermal-2}Linear thermal expansion in the tetragonal
plane measured along the $a$ axis and along the $\left[110\right]$
direction. The inset shows the corresponding thermal-expansion coefficient
for both directions. }

\end{figure}

In order to test the presence of a lattice distortion in $\mathrm{UAu_{2}Si_{2}}$
we measured the thermal expansion also along the $a$ axis and $\left[110\right]$\footnote{we use Miller indices notation for this direction for its self-explanatory
meaning, while for simplicity $a$ and $c$ are used for the $\left[100\right]$
and $\left[001\right]$ directions, respectively throughout the whole
paper} direction. The corresponding thermal expansions and the linear thermal-expansion
coefficients are plotted in Fig. \ref{fig:The-linear-thermal-2}. 

If the fourfold rotational symmetry of the tetragonal structure of
$\mathrm{UAu_{2}Si_{2}}$ in the UAFM phase can be broken similar
to the case of isostructural $\mathrm{URu_{2}Si_{2}}$ the crystal
structure itself is expected to have an orthorhombic distortion, lowering
the space symmetry. For the $I4/mmm$ space group, two subgroups $Fmmm$
and $Immm$ may have such orthorhombic distortions, but the $ab$
plane primitive vector direction is rotated by 45\textdegree{} (Ref.
\cite{Tonegawa2014}). This behavior is not seen in a macroscopic
sample in ambient condition owing to the formation of domains. A small
uniaxial pressure applied on multiple domains may change the domain
structure towards an orthorhombic monodomain state so that the distortion
may be indicated also macroscopically as suggested in case of $\mathrm{URu_{2}Si_{2}}$.\cite{Kambe2013a}
That is probably why we could observe the distortion in our dilatometer
which exerts a slight uniaxial pressure along the direction of measurement. 

The thermodynamic relation for second-order phase transitions known
as Ehrenfest relation allows us to estimate the pressure dependence
of the ordering temperature based on the jumps in the specific heat
$\Delta C$ and the thermal-expansion coefficient $\Delta\alpha$
at $T_{\mathrm{m}}$. It is defined as 

\begin{equation}
\frac{\mathrm{d}T_{\mathrm{m}}}{\mathrm{d}p}=\frac{\Delta\alpha V_{\mathrm{m}}}{\Delta C/T}\thinspace,\label{eq: Ehrenfest}
\end{equation}
and it can serve as an estimation for the hydrostatic-pressure dependence
using the volume thermal-expansion coefficient $\alpha_{v}$, or for
the uniaxial pressure dependences when the linear thermal-expansion
coefficients $\alpha_{a}$ or $\alpha_{c}$ are used. The jumps of
the thermal-expansion coefficients at $T_{\mathrm{m}}$ and the corresponding
hydrostatic-pressure and uniaxial-pressure derivatives of $T_{\mathrm{m}}$
predicted using Eq. (\ref{eq: Ehrenfest}) are shown in Table \ref{tab:The-jumps-of}.

\begin{table}
\caption{\label{tab:The-jumps-of}The jumps of the thermal-expansion coefficients
$\Delta\alpha_{i}$ at $T_{\mathrm{m}}$ and the corresponding hydrostatic-pressure
and uniaxial-pressure derivatives of $T_{\mathrm{m}}$ predicted using
Ehrenfest relation.}
\begin{centering}
\begin{tabular*}{8.6cm}{@{\extracolsep{\fill}}>{\centering}p{2cm}>{\centering}p{3.3cm}>{\raggedleft}p{1.65cm}@{\extracolsep{0pt}.}p{1.65cm}}
\toprule 
 & $\unit[\Delta\alpha_{i}]{\left(K^{-1}\right)}$ & \multicolumn{2}{p{3.3cm}}{$\unit[\frac{\mathrm{d}T_{\mathrm{m}}}{\mathrm{d}p}]{\left(K\,GPa^{-1}\right)}$}\tabularnewline
\midrule
\midrule 
$a$ & $2.69(8)\times10^{-5}$ & 2&7(1)\tabularnewline
\midrule 
$c$ & $-4.4(1)\times10^{-6}$  & -0&44(1)\tabularnewline
\midrule 
Volume & $4.9(1)\times10^{-5}$ & 4&9(1)\tabularnewline
\bottomrule
\end{tabular*}
\par\end{centering}
\end{table}

\subsection{Specific-heat, magnetization and magnetostriction measurements}

\begin{figure}
\begin{centering}
\includegraphics[scale=0.33]{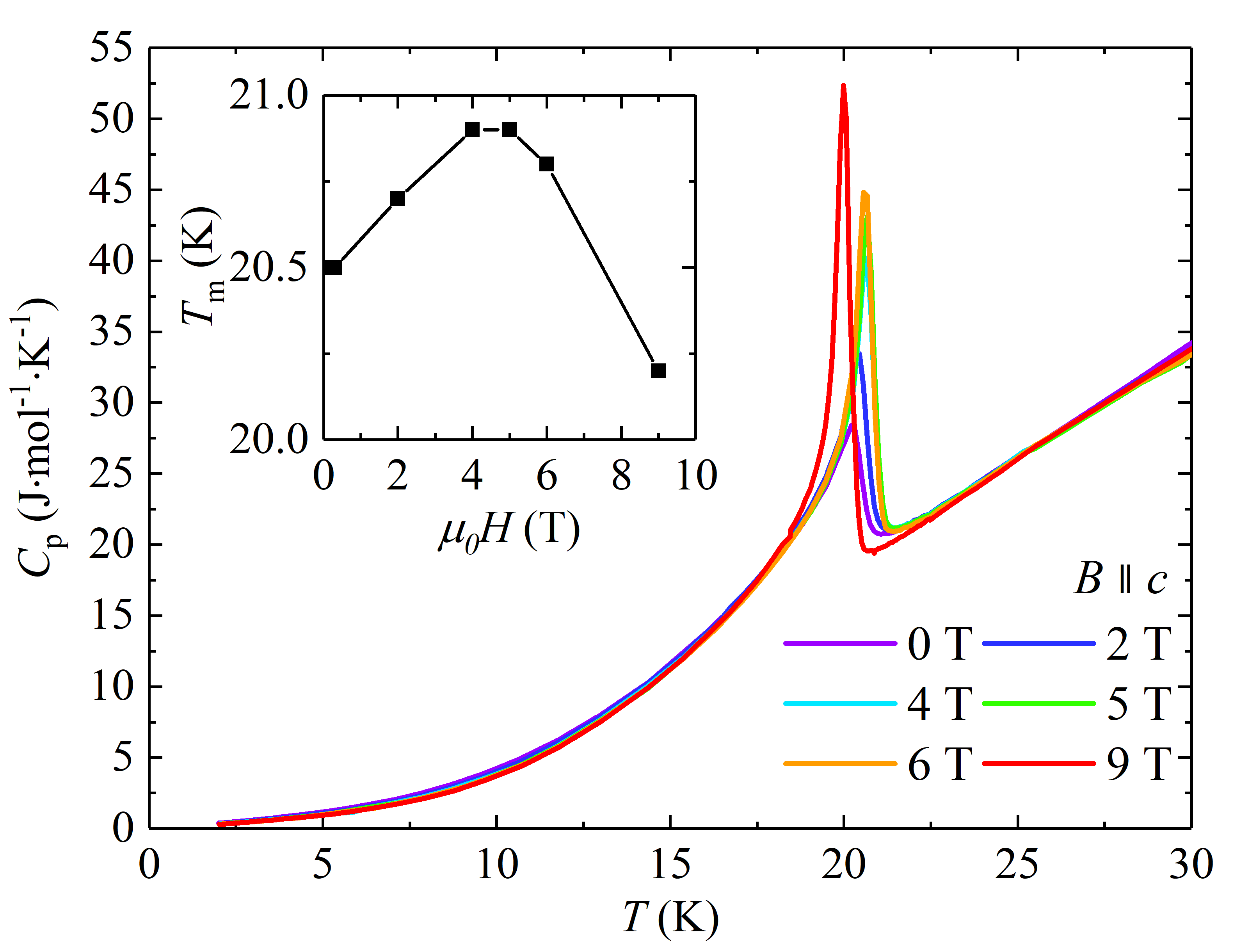}
\par\end{centering}
\caption{\label{fig:Temperature-dependence-of}Temperature dependence of specific
heat of $\mathrm{UAu_{2}Si_{2}}$ in magnetic fields applied along
the $c$ axis. Inset: Magnetic-field dependence of $T_{\mathrm{m}}$.}

\end{figure}

First we measured temperature dependences of specific-heat in various
magnetic fields applied along the $c$ axis. In Fig. \ref{fig:Temperature-dependence-of}
we can indeed see that $T_{\mathrm{m}}$ increases with magnetic field
up to $\sim\unit[4]{T}$ reaching the maximum value between $4$ and
$\unit[5]{T}$ and then decreasing with further increasing the field.

\begin{figure}
\begin{centering}
\includegraphics[scale=0.33]{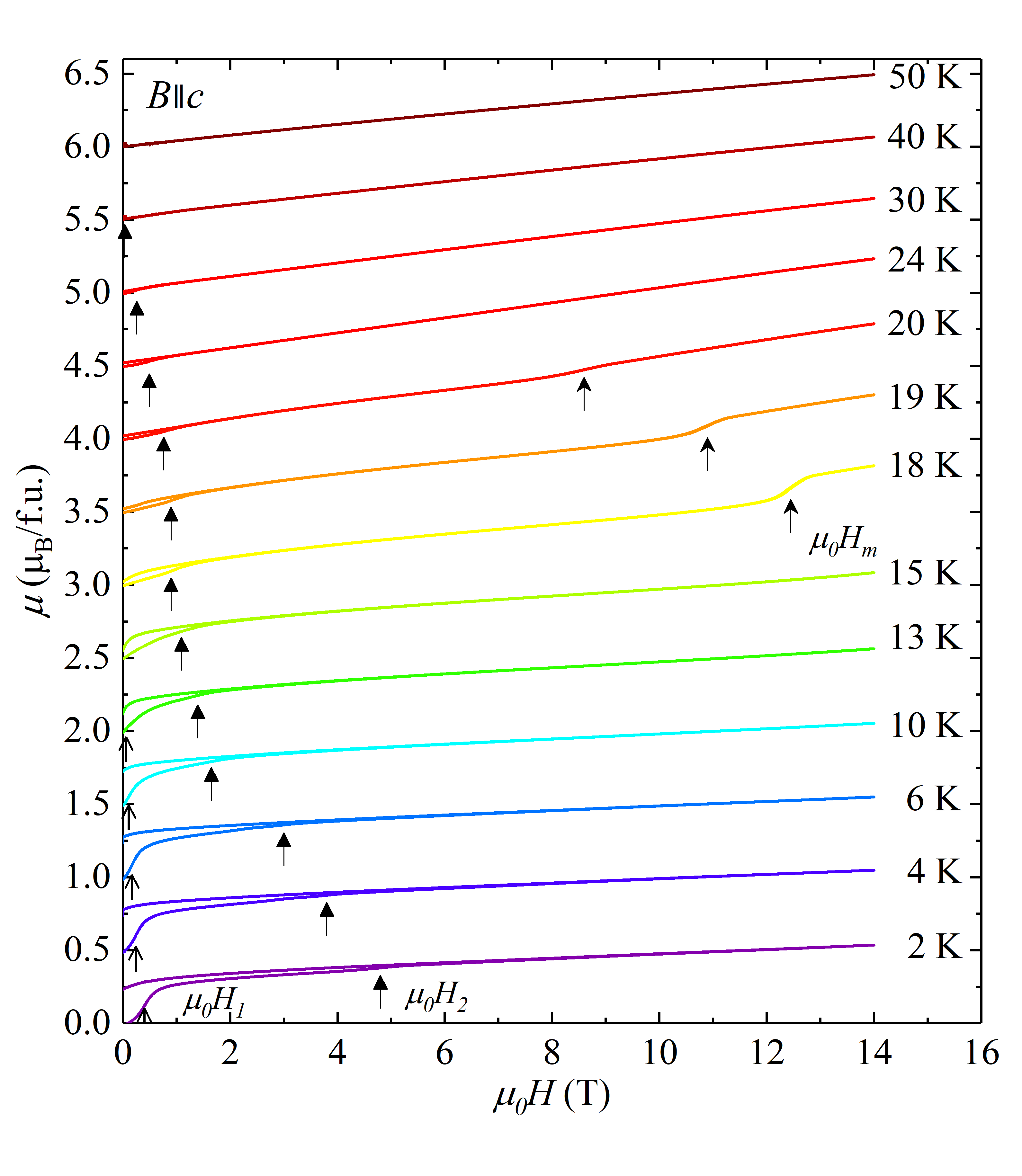}
\par\end{centering}
\caption{\label{fig:The-magnetization-curves}Magnetization curves measured
with the field applied along the $c$ axis up to $\unit[14]{T}$.
The curves are consecutively shifted by $\unit[0.5]{\mu_{\mathrm{B}}/f.u.}$
along the magnetization axis. The three types of arrows mark $\mu_{0}H_{1}$,
$\mu_{0}H_{2}$, and $\mu_{0}H_{m}$, respectively. }

\end{figure}

\begin{figure}
\begin{centering}
\includegraphics[scale=0.33]{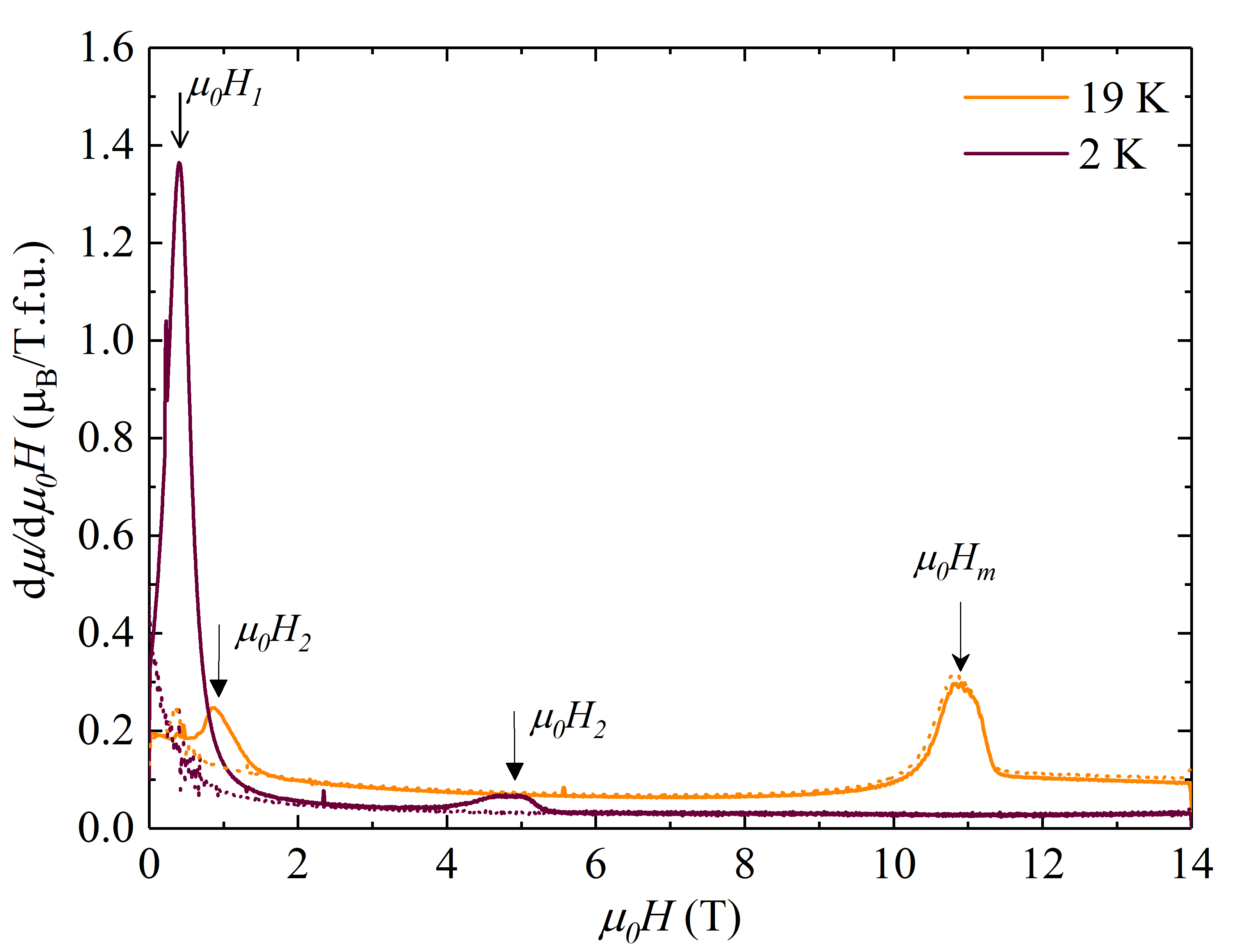}
\par\end{centering}
\caption{\label{fig:The-example-of}Field derivatives of the magnetization
data at $2$ and $\unit[19]{K}$ depicting the determination of $\mu_{0}H_{1}$,
$\mu_{0}H_{2}$, and $\mu_{0}H_{m}$. The solid lines correspond to
field-up and dotted lines to field-down sweeps. }

\end{figure}

The magnetization was measured at various temperatures from $2$ to
$\unit[50]{K}$ as function of magnetic field up to $\unit[14]{T}$
applied along the $c$ axis. This was followed by high-field measurements
in pulsed magnetic fields up to $\sim\unit[58]{T}$. The results are
shown in Fig. \ref{fig:The-magnetization-curves}. The magnetization
data up to $\unit[13]{K}$ show a low-field inflection point in the
hysteretic part of the curve labeled as $\mu_{0}H_{1}$. Another step-like
feature is present at higher field and is labeled as $\mu_{0}H_{2}$.
Unlike the low-field transition, the $\mu_{0}H_{2}$ anomaly can be
traced to temperatures above $T_{\mathrm{m}}$ and is clearly distinguishable
even at $\unit[40]{K}$. We have discussed this behavior in Ref. \cite{Tabata2016},
especially in the parts connected with Figures 11, 18-20. Two components
apparently coexist at temperatures below $T_{\mathrm{m}}$: a) a very
weak ferromagnetic one which emerges around $\unit[50]{K}$ with cooling;
its coercive field $\mu_{0}H_{2}$ increases with decreasing temperature
and exceeds $\unit[4]{T}$ at $\unit[2]{K}$, b) an uncompensated
AFM one (UAFM) with a considerably larger spontaneous magnetization
but a much smaller coercive field $\mu_{0}H_{1}$. The magnetization
isotherms in the vicinity of $T_{\mathrm{m}}$ show another field-induced
step-like transition at higher fields. Similar transitions were observed
in our previous work on a different single crystal.\cite{Tabata2016}
There, the anomalies labeled as $\mu_{0}H_{m}$ were, however, much
less pronounced probably due to lower crystal quality. The values
of characteristic fields at different temperatures have been determined
using plots of $\mathrm{d}\mu/\mathrm{d}\mu_{0}H$ vs. $\mu_{0}H$
shown in Fig. \ref{fig:The-example-of} for results at $2$ and $\unit[19]{K}$.

\begin{figure}
\begin{centering}
\includegraphics[scale=0.33]{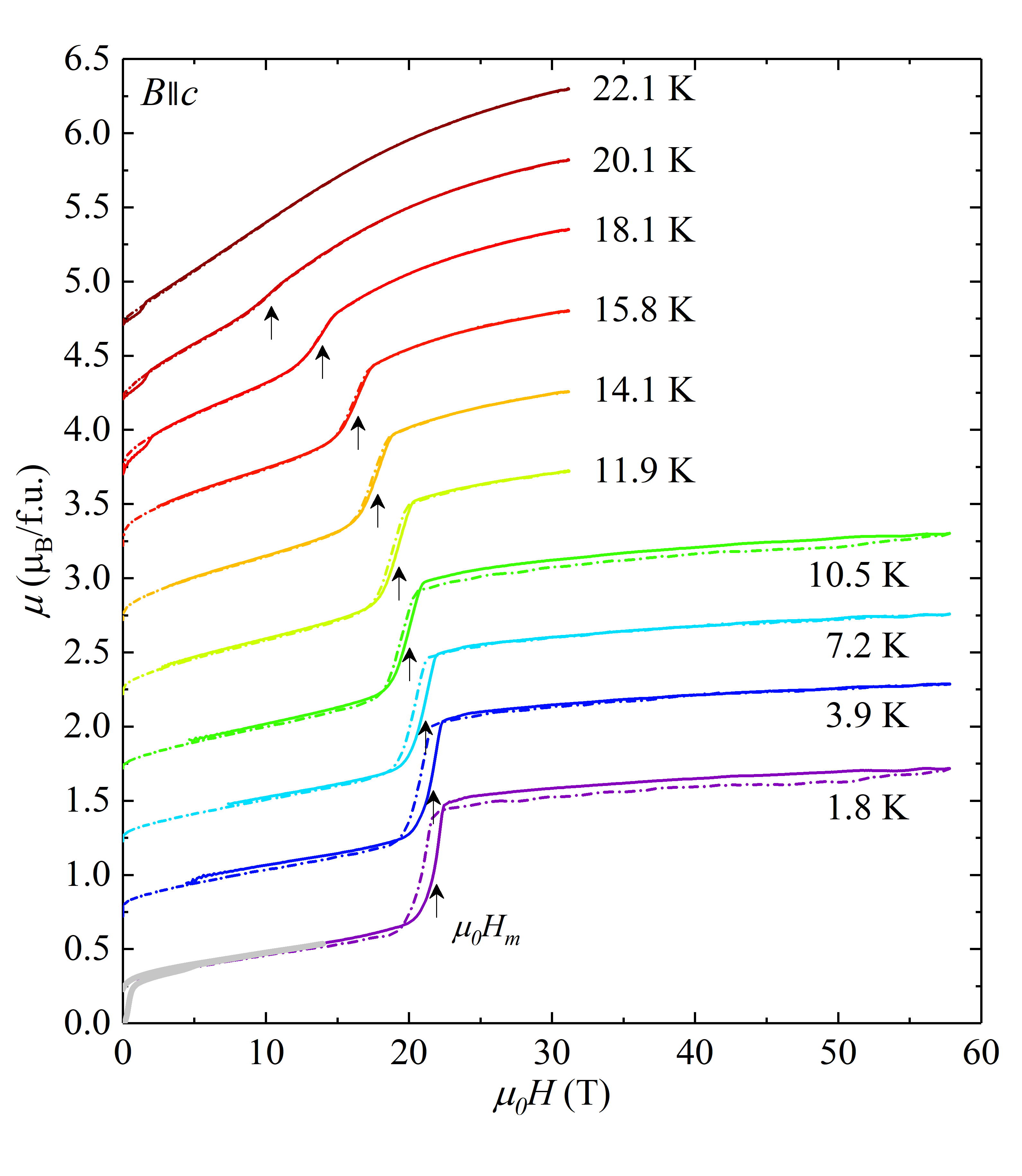}
\par\end{centering}
\caption{\label{fig:The-magnetization-curves-1}Magnetization measured in pulsed
fields applied along the $c$ axis up to $\approx\unit[58]{T}$. The
curves are consecutively vertically shifted by $\unit[0.5]{\mu_{\mathrm{B}}/f.u.}$.
The solid lines correspond to field-up and dash-dotted lines to field-down
sweeps. The arrows mark the $\mu_{0}H_{\mathrm{m}}$ transitions.
The gray curve overlapping data at $\unit[1.8]{K}$ are the static-field
data at $\unit[2]{K}$ up to $\unit[14]{T}$ shown in Fig. \ref{fig:The-magnetization-curves}.}

\end{figure}

Further magnetization measurements in pulsed fields were performed
to track the metamagnetic transition $\mu_{0}H_{m}$. The measured
magnetization was scaled using the static-field data obtained at $\unit[2]{K}$
and a small linear background was subtracted to give the correct absolute
values. Results of the high-field measurement are plotted in Fig.
\ref{fig:The-magnetization-curves-1}. A clear metamagnetic transition
at $\mu_{0}H_{m}$ is visible on all measured isotherms up to $\unit[20.1]{K}$.
The metamagnetic transition is most probably of spin-flip type\cite{Stryjewski1977}
as one would expect for the system with large magnetocrystalline anisotropy.
The low-temperature curves also show significant hystereses ($\Delta\mu_{0}H_{m}$)
between the up and down sweeps. The hysteresis narrows with increasing
temperature and disappears around $\unit[16]{K}$ when $\mu_{0}H_{\mathrm{m}}\sim\unit[16]{T}$.
The temperature dependence of $\Delta\mu_{0}H_{m}$ is shown in Fig.
\ref{fig:a)-The-magnetic}. 

\begin{figure}
\begin{centering}
\includegraphics[scale=0.33]{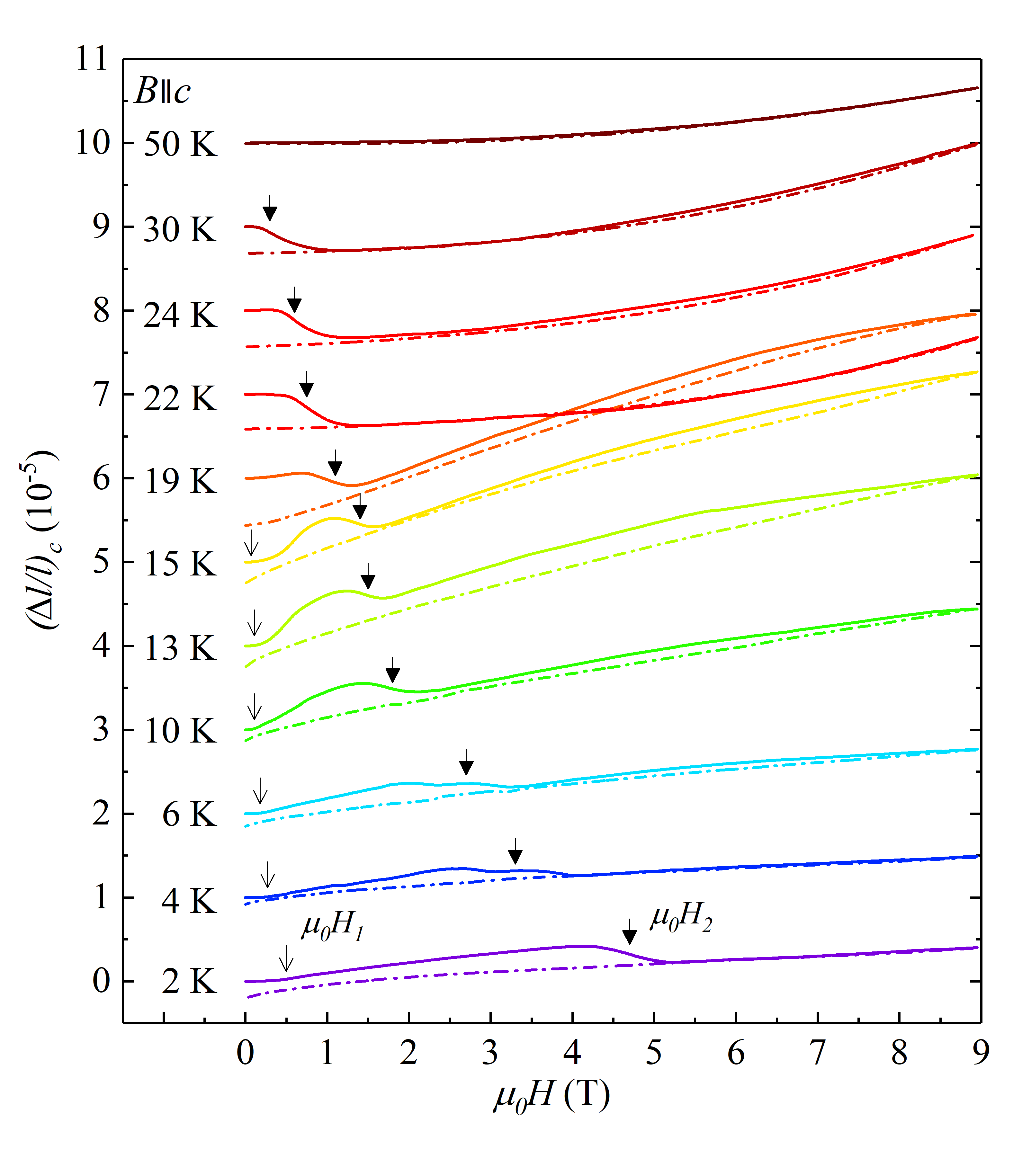}
\par\end{centering}
\caption{\label{fig:The-longitudinal-magnetostrictio}Longitudinal magnetostriction
measured along the $c$ axis. The solid lines correspond to field-up
sweeps and dash-dotted lines to field-down sweeps. The curves are
vertically shifted for better clarity.}

\end{figure}

The longitudinal magnetostriction was measured at various temperatures
with magnetic fields up to $\unit[9]{T}$ applied along the tetragonal
$c$ axis, see Fig. \ref{fig:The-longitudinal-magnetostrictio}.

\begin{figure}
\begin{centering}
\includegraphics[scale=0.33]{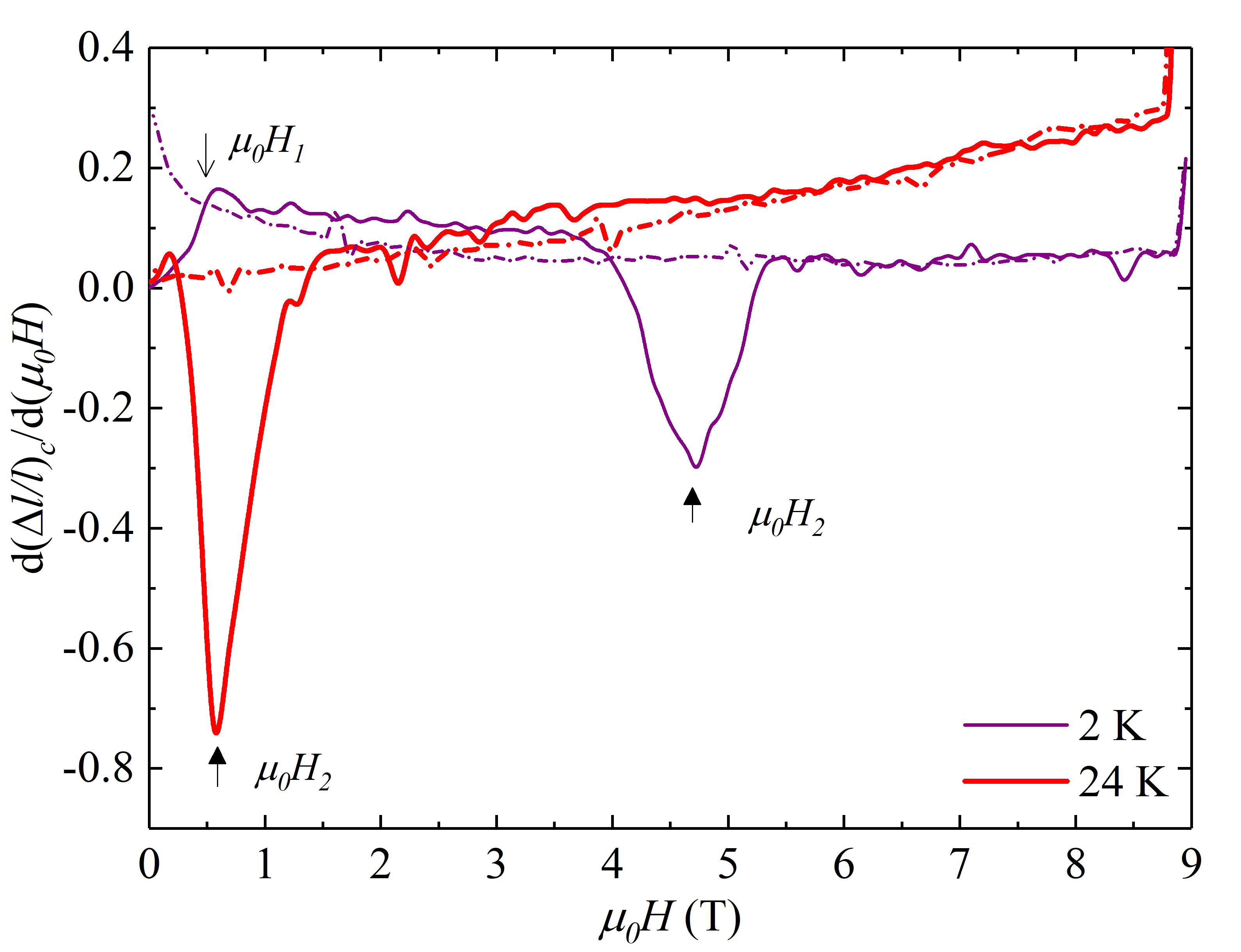}
\par\end{centering}
\caption{\label{fig:The-example-of-1}Field derivatives of the magnetostriction
data at $2$ and $\unit[24]{K}$ showing the determination of $\mu_{0}H_{1}$
and $\mu_{0}H_{2}$. The solid lines correspond to field-up scans
and dash-dotted lines to field-down sweeps.}

\end{figure}

Fig. \ref{fig:The-example-of-1} shows how the coercive fields $\mu_{0}H_{1}$
and $\mu_{0}H_{2}$ were determined from magnetostriction data.

\begin{figure}
\begin{centering}
\includegraphics[scale=0.33]{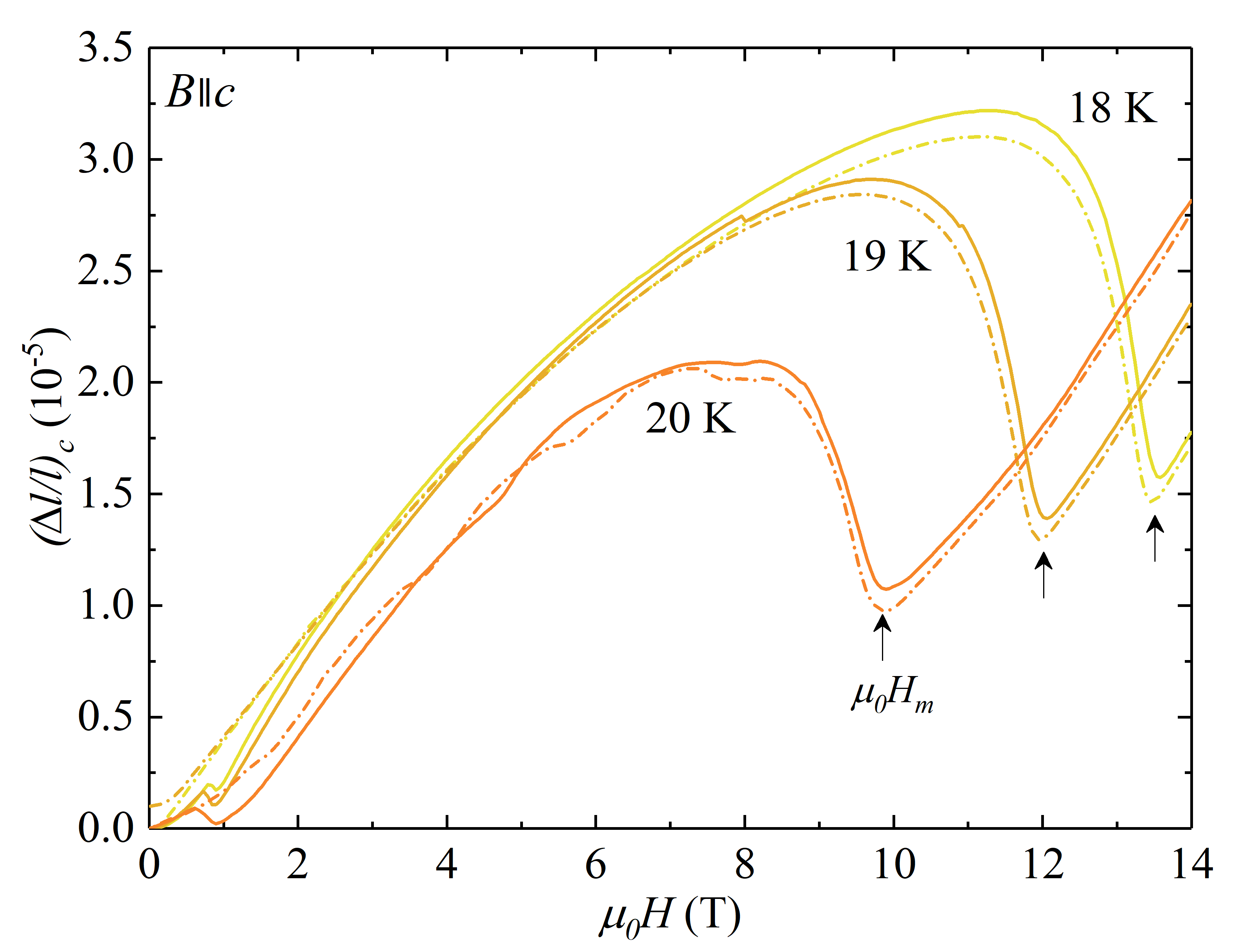}
\par\end{centering}
\caption{\label{fig:The-magnetostriction-close}Longitudinal magnetostriction
measured with the fields up to the $\unit[14]{T}$ applied along the
$c$ axis.}

\end{figure}

There is a significant change of the shape of the magnetostriction
curves when $\mathrm{UAu_{2}Si_{2}}$ crosses $T_{\mathrm{m}}=\unit[20.5]{K}$,
going from a concave to a convex curvature. We have consequently conducted
further measurements of the longitudinal magnetostriction along the
$c$ axis up to $\unit[14]{T}$ at selected temperatures close to
$T_{\mathrm{m}}$. These isotherms show the high-field anomaly as
a pronounced sharp drop at $\mu_{0}H_{m}$ (see Fig. \ref{fig:The-magnetostriction-close}),
as determined from the magnetization measurements. 

\begin{figure}
\begin{centering}
\includegraphics[scale=0.33]{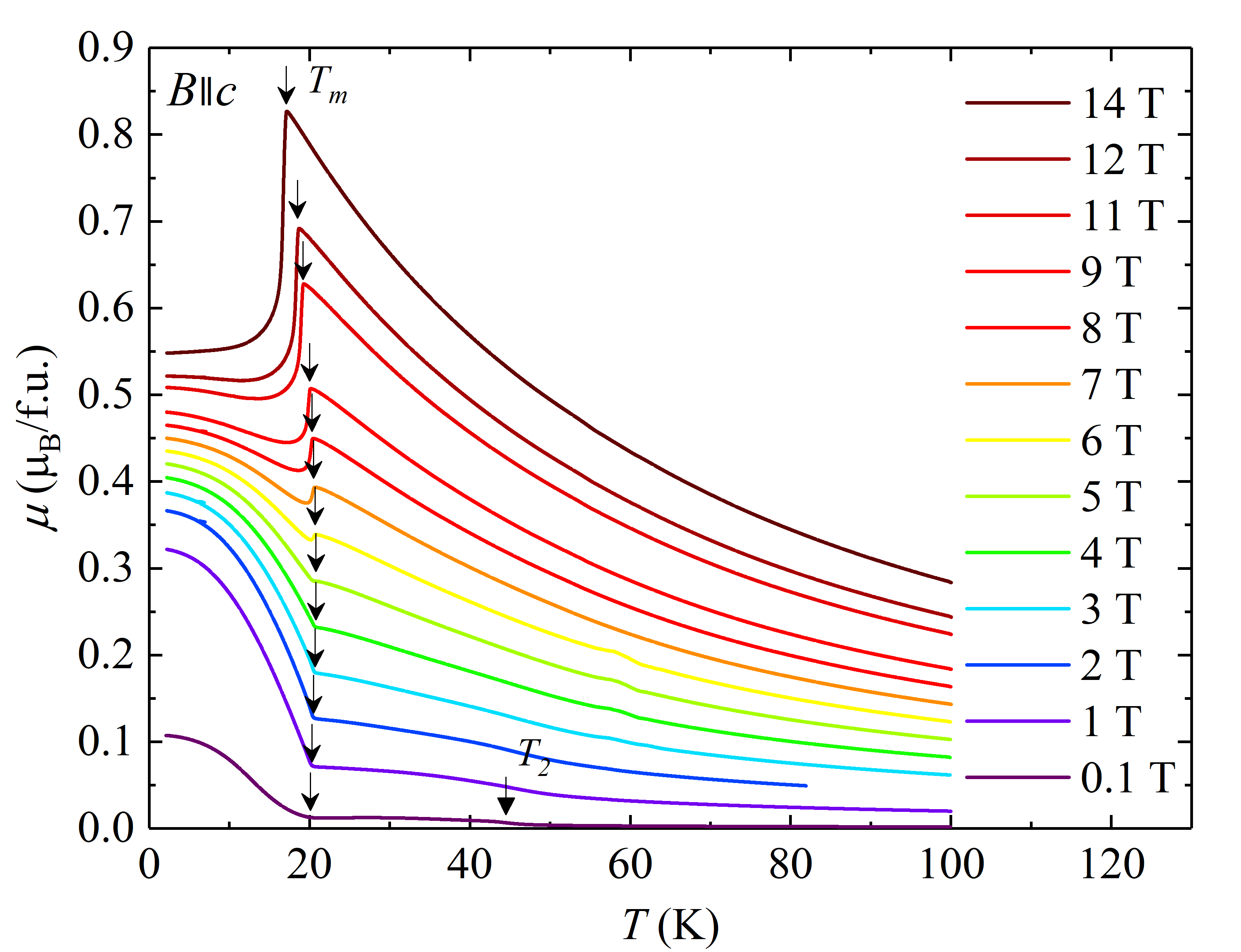}
\par\end{centering}
\caption{\label{fig:The-temperature-dependence}Temperature dependence of the
magnetization along the $c$ axis. The arrows mark the $T_{\mathrm{m}}$
and $T_{\mathrm{2}}$ transitions.}
\end{figure}

\begin{figure}
\begin{centering}
\includegraphics[scale=0.33]{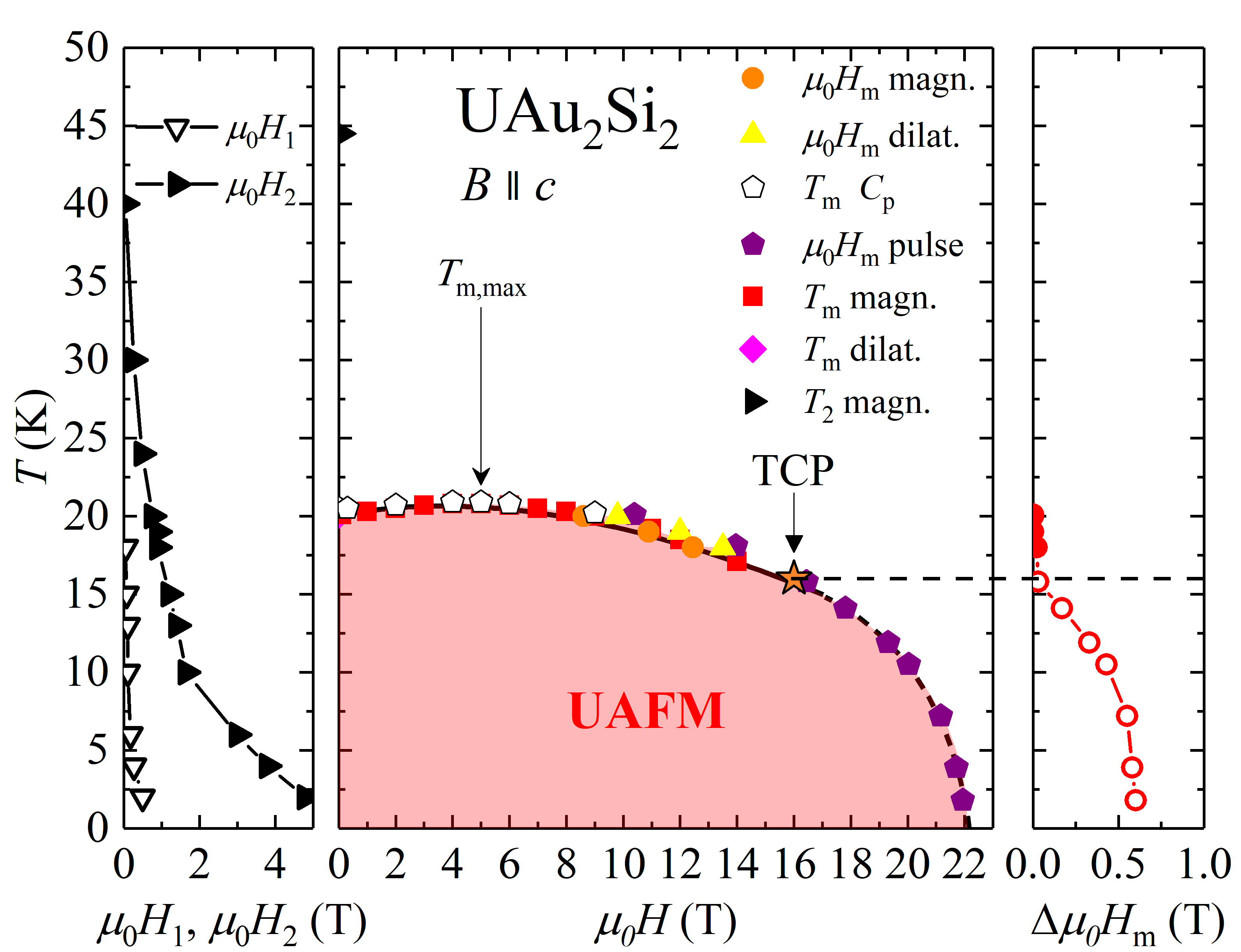}
\par\end{centering}
\caption{\label{fig:a)-The-magnetic}(middle panel) Magnetic phase diagram
of $\mathrm{UAu_{2}Si_{2}}$ constructed using the results of the
specific-heat - $C_{\mathrm{p}}$, magnetization and dilatometric
(thermal expansion, magnetostriction) measurements. The star marks
the tricritical point - TCP. (left panel) Temperature dependences
of $\mu_{0}H_{1}$, and $\mu_{0}H_{2}$. (right panel) Temperature
dependence of the hysteresis of the metamagnetic transition $\mu_{0}H_{m}$
resulting from the pulsed-field measurements - open symbols and from
the static-field measurements - full symbols.}

\end{figure}

We have also measured the temperature dependence of the magnetization
in various fields up to $\unit[14]{T}$ applied along the $c$ axis.
These data agree with our previous results obtained on a different
single crystal.\cite{Tabata2016} We can clearly see the anomaly labeled
as $T_{\mathrm{m}}$, determined from the upturn of the magnetization
for curves below $\unit[5]{T}$ and from the peak at higher fields.
There is also another transition marked as $T_{\mathrm{2}}$ that
can be distinguished only in the low-field data at $\unit[0.1]{T}$
(see Fig. \ref{fig:The-temperature-dependence}).

The whole set of anomalies observed in the magnetization, magnetostriction
and thermal-expansion measurements allows us to construct the magnetic
phase diagram as plotted in Fig. \ref{fig:a)-The-magnetic} together
with the temperature dependence of the hystereses of the $\mu_{0}H_{1}$,
$\mu_{0}H_{2}$ and $\mu_{0}H_{m}$.

\subsection{\label{subsec:Hydrostatic-pressure-study}Hydrostatic pressure study}

\begin{figure}
\begin{centering}
\includegraphics[scale=0.33]{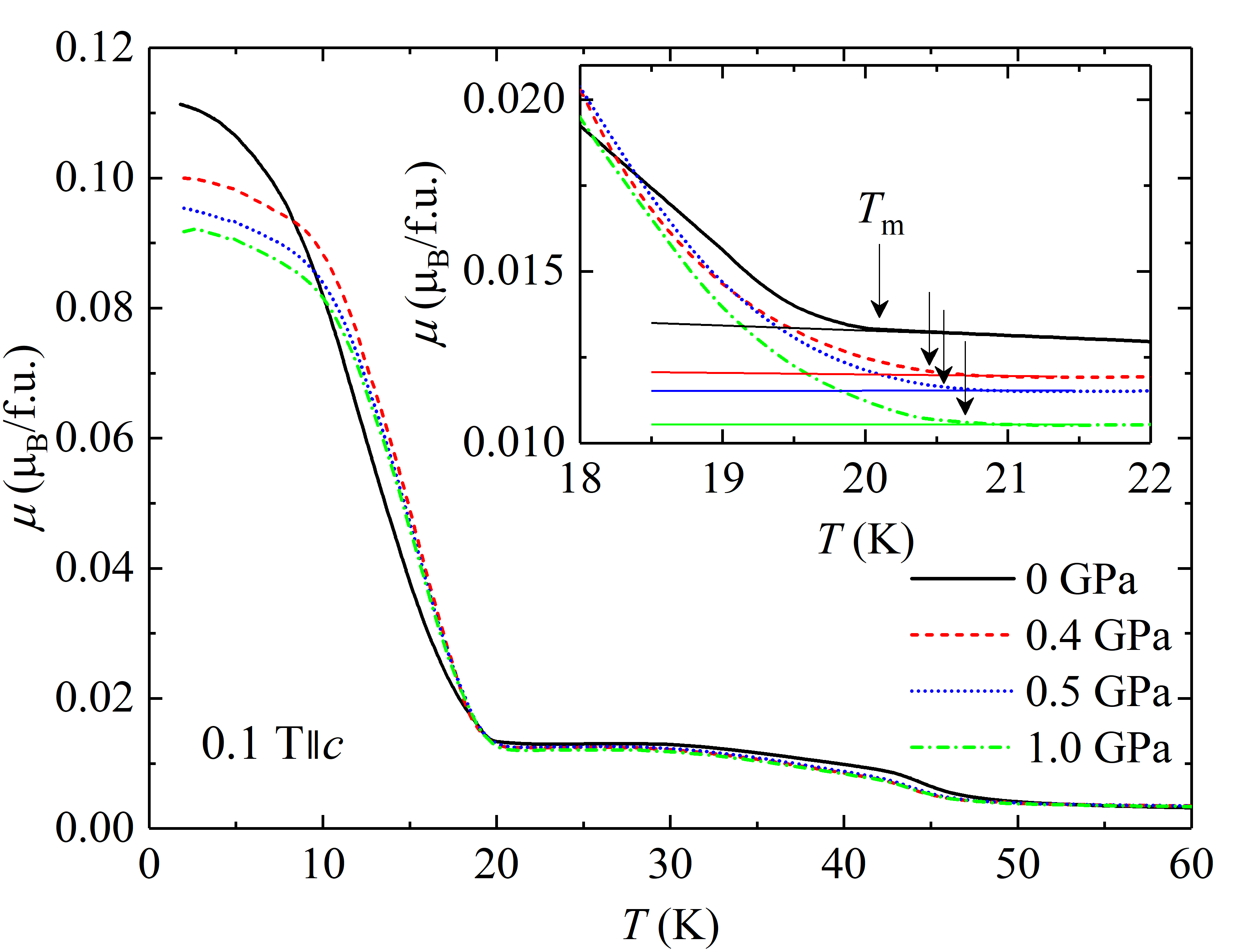}
\par\end{centering}
\caption{\label{fig:Results-of-the}Magnetization in a magnetic field of $\unit[0.1]{T}$
applied along the $c$ axis under various hydrostatic pressures up
to $\unit[1.0]{GPa}$. The inset shows the region near $T_{\mathrm{m}}$
marked by the arrows. The curves in the inset are vertically shifted
for clarity. }

\end{figure}

As our calculations using the Ehrenfest relation predict a rather
dramatic positive effect of hydrostatic pressure on the ordering temperature
($\approx\unit[4.9(1)]{K\,GPa^{-1}}$) we wanted to verify this hypothesis.
For that purpose, we measured the magnetization in a field of $\unit[0.1]{T}$
applied along the $c$ axis under hydrostatic pressures up to $\sim\unit[1]{GPa}$
(see Fig. \ref{fig:Results-of-the}). The measured data were corrected
for the diamagnetic contribution of the pressure cell. The shape of
the ambient pressure curve differs from those obtained under pressure.
This can be an effect of a slightly different orientation of the sample
in the pressure cell. Contrary to our prediction, we observe only
a small shift of the transition temperature $T_{\mathrm{m}}$ with
applied pressure (see inset of Fig. \ref{fig:Results-of-the}). The
transition temperature $T_{\mathrm{m}}$ again is defined by the upturn
of the magnetization curve. The resulting small ratio of the pressure
change of the ordering temperature is $\mathrm{d}T_{\mathrm{m}}/\mathrm{d}p\approx\unit[0.6(1)]{K\,GPa^{-1}}$.
A larger effect is visible in the reduction of the spontaneous magnetic
moment $\mu_{\mathrm{spont}}$ with the slope $\mathrm{d}\mu_{\mathrm{spont}}/\mathrm{d}p\approx\unit[-0.019(6)]{\mu_{\mathrm{B}}/\left(f.u.\,GPa\right)}$. 

\begin{figure}
\begin{centering}
\includegraphics[scale=0.33]{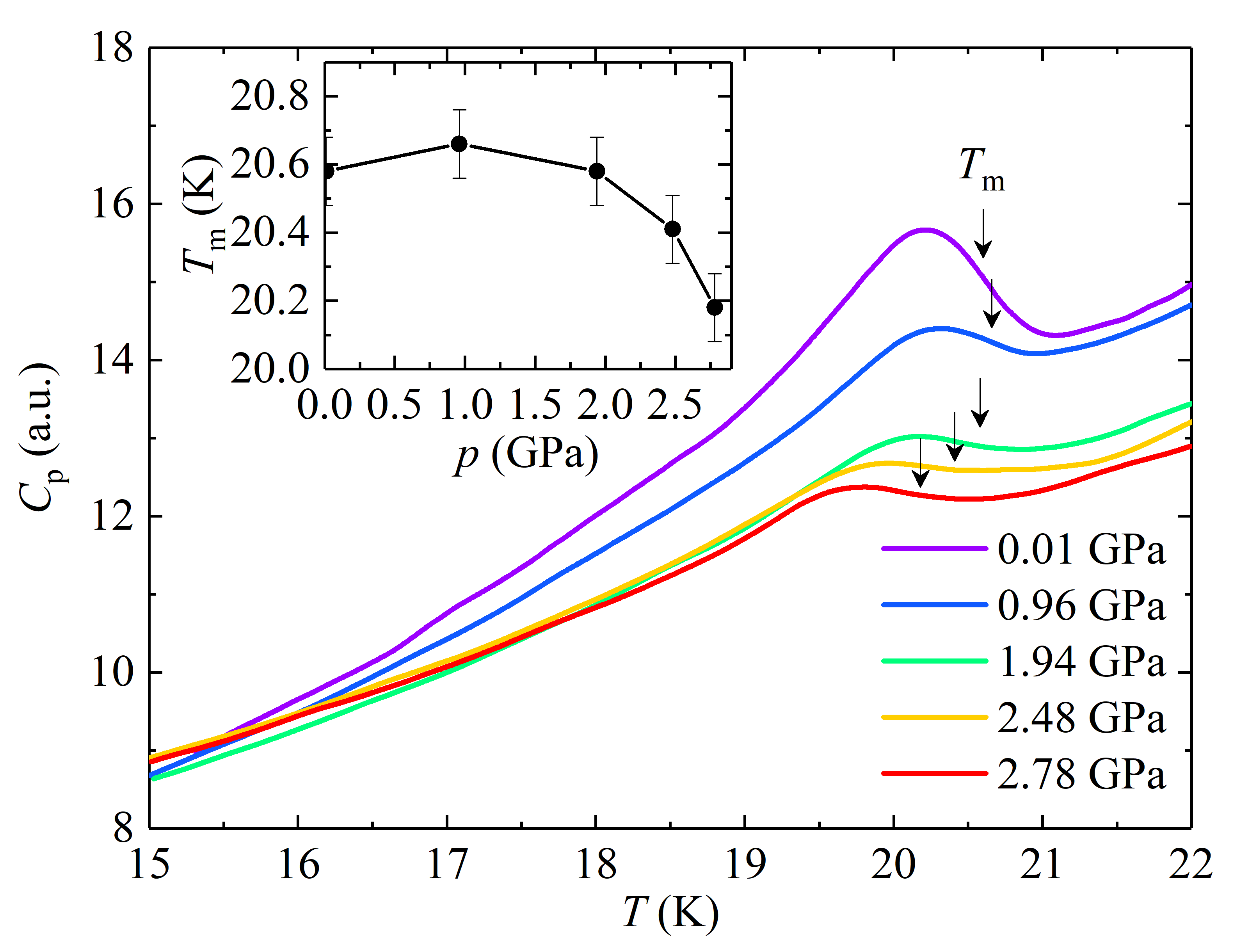}
\par\end{centering}
\caption{\label{fig:Temperature-dependence-of-1}Temperature dependence of
specific heat of $\mathrm{UAu_{2}Si_{2}}$ under hydrostatic pressure.
Inset: Pressure dependence of $T_{\mathrm{m}}$.}
\end{figure}

Temperature dependence of specific heat measured under hydrostatic
pressure up to $\unit[2.79]{GPa}$ confirmed that $T_{\mathrm{m}}$
slightly increases with applying pressure of $\unit[1]{GPa}$ and
then decreases at gradually increasing rate with higher pressure (see
Fig. \ref{fig:Temperature-dependence-of-1}). The total change of
$T_{\mathrm{m}}$ between ambient pressure and $\unit[2.79]{GPa}$
amounts only $\unit[-0.4]{K}$, i.e. $\sim-2\%$ . In any case the
results of measurements of pressure influence on $T_{\mathrm{m}}$
strongly contradict the predictions from Ehrenfest relation. This
contradiction obviously requires further studies. At this stage we
can only speculate that the reasons can be in a complex hierarchy
of exchange interactions. 

\section{Discussion}

Our thermal-expansion measurements on a $\mathrm{UAu_{2}Si_{2}}$
single crystal revealed a large anisotropy which is mainly due to
magnetic contributions. The ordered state below $T_{\mathrm{m}}$
is connected with a dramatic lattice contraction in the basal plane
($a$ axis). This together with the relatively small expansion of
the $c$ axis leads to the ground state volume collapse (Fig. \ref{fig:The-linear-thermal}).
This behavior strongly resembles the case of the isostructural heavy-fermion
compound $\mathrm{URu_{2}Si_{2}}$ entering the hidden-order state
at $\unit[17.5]{K}$.\cite{deVisser1986} The linear thermal-expansion
coefficient for the $a$ axis consequently exhibits a sharp positive
anomaly at the ordering temperature whereas the $c$-axis anomaly
is negative and less pronounced (Fig. \ref{fig:The-linear-thermal-1}).
The volume thermal-expansion coefficient of $\mathrm{URu_{2}Si_{2}}$
also exhibits a sharp and positive peak at $\unit[17.5]{K}$ that
can be translated to a volume decrease in the ground state. A very
similar behavior for both compounds can be found in the temperature
dependence of the $c/a$ ratio. Both materials show an upturn below
the ordering temperature, accenting the contraction in the basal plane,
and nearly linear temperature dependence at higher temperatures up
to $\unit[40]{K}$. Nevertheless, the $c/a$ ratio of $\mathrm{URu_{2}Si_{2}}$
has a pronounced minimum around $\unit[60]{K}$, that is not visible
for $\mathrm{UAu_{2}Si_{2}}$ which shows a linear temperature dependence
up to $\unit[100]{K}$. Even though the overall character of the thermal
expansion of $\mathrm{URu_{2}Si_{2}}$ and $\mathrm{UAu_{2}Si_{2}}$
is qualitatively similar, it does differ quantitatively. The step
in the volume thermal-expansion coefficient $\alpha_{v}^{*}=\left(\alpha_{a}+\alpha_{b}+\alpha_{c}\right)/3$
is $\sim\unit[2.5\times10^{-6}]{K^{-1}}$ for $\mathrm{URu_{2}Si_{2}}$
\cite{deVisser1986} and almost an order of magnitude larger ($\sim\unit[1.6\times10^{-5}]{K^{-1}}$)
for $\mathrm{UAu_{2}Si_{2}}$. It is believed, that anomalies in the
thermal-expansion coefficient of the order of $10^{-4}$, $10^{-5}$
can be connected with a structural transitions\cite{deVisser1986}
as in the case of $\mathrm{UPd_{3}}$.\cite{Ott1980} In that sense,
$\mathrm{URu_{2}Si_{2}}$ does not evidence a structural change in
the hidden-order state. Nevertheless, there is a list of studies which
suggest the breaking of the fourfold rotational symmetry of the tetragonal
$c$ axis,\cite{Okazaki2011,Shibauchi2014,Tonegawa2012,Tonegawa2013,Kambe2013,Riggs2015}
whereas the high-resolution x-ray backscattering\cite{Tabata2014}
and thermal–expansion data\cite{deVisser1986,Kuwahara1997} do not
confirm this. However, lattice-symmetry breaking from the fourfold
tetragonal to twofold orthorhombic structure was unambiguously observed
by high-resolution synchrotron x-ray diffraction measurements in zero
field.\cite{Tonegawa2014} The fact that this distortion is observed
only in ultra-pure samples may explain the long list of more or less
unsuccessful attempts to observe this. 

As the thermal-expansion coefficients of $\mathrm{UAu_{2}Si_{2}}$
are even one order of magnitude larger (i.e., $\sim10^{-5}$) the
possibility of some kind of lattice distortion should be seriously
considered. Our thermal-expansion measurements show an anisotropic
expansion in the basal plane breaking the fourfold symmetry along
the $c$ axis. The body-centered room-temperature tetragonal structure
of $\mathrm{UAu_{2}Si_{2}}$ belongs to the $I4/mmm$ space group.
It has 15 maximal non-isomorphic subgroups and only two of them have
no fourfold symmetry along the $c$ axis. These are the orthorhombic
$Fmmm$ and $Immm$ space groups. The same space groups were also
considered in the synchrotron x-ray diffraction study of $\mathrm{URu_{2}Si_{2}}$.\cite{Tonegawa2014}
The $Fmmm$ space group was found to describe the system in the hidden-order
state. 

The measurement of thermal expansion, as a macroscopic quantity, is
not sufficient to properly describe the space group of the distorted
structure, even though it is more sensitive to detect distortions
than diffraction studies. In that sense, high-resolution x-ray diffraction
experiments are needed to resolve the structure of $\mathrm{UAu_{2}Si_{2}}$
in the ordered state. Our results from an ultrasonic study show a
Curie-type softening in the transverse $\left(C_{11}-C_{12}\right)/2$
mode toward $T_{\mathrm{m}}$, that could also point to orthorhombic
distortion at $T_{\mathrm{m}}$.\cite{Yanagisawa} In the analogy
with the $\mathrm{URu_{2}Si_{2}}$, there was also observed softening
of the same mode suggesting that the $\Gamma_{3\mathrm{g}}\left(\mathrm{B_{1g}}\right)$
-type lattice instability is innate in these systems.\cite{Kuwahara1997,Yanagisawa2012}

The large ground-state volume collapse of $\mathrm{UAu_{2}Si_{2}}$
indicates initial positive pressure dependence of the ordering temperature
of $\mathrm{d}T_{\mathrm{m}}/\mathrm{d}p\approx\unit[4.9(1)]{K\,GPa^{-1}}$,
according to the Ehrenfest relation. Uniaxial pressure applied along
the $a$ axis should have a positive effect as well ($\approx\unit[2.7(1)]{K\,GPa^{-1}}$).
On the other hand, uniaxial pressure along the tetragonal $c$ axis
should lower $T_{\mathrm{m}}$ at a rate of $\approx\unit[-0.44(1)]{K\,GPa^{-1}}$.
These findings qualitatively agree with the experimentally confirmed
behavior of $\mathrm{URu_{2}Si_{2}}$ where the predicted pressure
dependences are approximately eight times smaller. Ehrenfest-relation
estimates give a pressure dependence of the hidden-order transition
of $\unit[1.4]{K\,GPa^{-1}}$(Ref. \cite{deVisser1986}) and high-pressure
resistivity measurements show an experimental initial slope of $\unit[1.01]{K\,GPa^{-1}}$
(Ref. \cite{Hassinger2008}). The estimated pressure changes of the
ordering temperature of the hidden order of $\mathrm{URu_{2}Si_{2}}$
and the UAFM state of $\mathrm{UAu_{2}Si_{2}}$ are largely different.
A similar dramatic change of the $\mathrm{d}T/\mathrm{d}p$ values
was observed for the $\mathrm{U\left(Ru,Fe\right)_{2}Si_{2}}$ system,
where doping of Fe leads to a change of the hidden order to “large-moment
antiferromagnetism”.\cite{Ran2016} However, our measurement of the
magnetization under hydrostatic pressure up to $\unit[1.0]{GPa}$
show only a weak pressure dependence of $T_{\mathrm{m}}$ (Fig. \ref{fig:Results-of-the}).
The estimated slope is $\mathrm{d}T_{\mathrm{m}}/\mathrm{d}p\approx\unit[0.6(1)]{K\,GPa^{-1}}$.
We also observed the lowering of the spontaneous magnetization with
increasing pressure $\mathrm{d}\mu_{\mathrm{spont}}/\mathrm{d}p\approx\unit[-0.019(6)]{\mu_{\mathrm{B}}/\left(f.u.\,GPa\right)}$.
The specific-heat measurement under hydrostatic pressure up to $\unit[2.79]{GPa}$
revealed small initial increase of $T_{\mathrm{m}}$ up to $\unit[1]{GPa}$
followed by suppression for higher pressure. The observed inconsistency
with the expected trend from the Ehrenfest relation is rather unexpected.
It may be caused by a structural distortion that takes place at $T_{\mathrm{m}}$.
In that case Eq. (\ref{eq:volume}) is not valid and the real volume
change can be different, i.e., possibly smaller. Another question
is the applicability of the Ehrenfest relation itself. Although it
is widely and successfully used to characterize the pressure dependence
(both positive and negative) of AFM\cite{Neumeier2001,Schmiedeshoff2006}
and FM\cite{Gasparini2010,Sakarya2003,Neumeier2001} second-order
phase transitions, it may strictly be applied only for the superconducting
transitions.\cite{Pippard1964} And even for some superconductors
the predicted pressure dependence determined by use of the Ehrenfest
relation differs from the experimental findings, such as in the case
of $\mathrm{PuCoGa_{5}}$\cite{Eloirdi2017} by an order of magnitude
or even by sign in the layered iron-based superconductors of the $\mathrm{Ba(Fe_{1-x}Co_{x})_{2}As_{2}}$
series.\cite{daLuz2009} Both of these systems exhibit very anisotropic
thermal-expansion coefficients, similar as for $\mathrm{UAu_{2}Si_{2}}$.

The magnetization isotherms (Fig. \ref{fig:The-magnetization-curves}
and \ref{fig:The-magnetization-curves-1}) clearly and reproducibly
show anomalies at $\mu_{0}H_{1}$, $\mu_{0}H_{2}$, and $\mu_{0}H_{m}$
in line with our previous study on a different single crystal. However,
we now found a much sharper character of the step-like transition
at $\mu_{0}H_{m}$. We suggest, that the deviation from the linear
dependence of the magnetization, that was marked as $\mu_{0}H_{m}$
in our previous work,\cite{Tabata2016} is a low-field sign of the
step-like transition which takes place at higher fields. We have previously
not observed this transition, possibly due to a lower crystal quality
or slightly improper orientation of the $c$ axis with respect to
the applied field. 

Our magnetostriction measurements (Fig. \ref{fig:The-longitudinal-magnetostrictio}
and \ref{fig:The-magnetostriction-close}) with field applied along
the $c$ axis reproduce the anomalies observed in the magnetization
data. The transition at $T_{2}$ is reflected in the thermal-expansion
data only by a small slope change in $\alpha_{a}$, $\alpha_{v}$
and $\alpha_{c/a}$ around $\unit[50]{K}$ (Fig. \ref{fig:The-linear-thermal-1}).
The size of the relative length change of the $c$ axis at the $\mu_{0}H_{2}$
anomaly, is of the order of $10^{-6}$. This provides an evidence
of its bulk character, which can be traced up to $\sim\unit[40]{K}$
as in the magnetization data. A larger relative length contraction
($\sim10^{-5}$) takes place at $\mu_{0}H_{m}$. It resembles the
$c$-axis contraction at the field-induced phase transition of $\mathrm{URu_{2}Si_{2}}$.\cite{Correa2012}

Our high-field magnetization measurements show that the temperature
dependence of the hysteresis of the $\mu_{0}H_{m}$ transition vanishes
around $\unit[16]{K}$ (and $\unit[16]{T}$) where the transition
changes its character from step like to continuous. This is attributed
to the change of the order of the phase transition from first order
(in higher fields) to second order (in lower fields). Such a point
is usually referred to as a tricritical point (TCP).\cite{Stryjewski1977}
We marked this point by a star in the phase diagram in Fig. \ref{fig:a)-The-magnetic}.
Similar tricritical points have recently been reported in the uranium-based
antiferromagnets $\mathrm{USb_{2}}$\cite{Stillwell2017} and $\mathrm{UN}$.\cite{Shrestha2017}
The critical field, where the transition temperature $T_{\mathrm{m}}$
is suppressed to $\unit[0]{K}$ in $\mathrm{UAu_{2}Si_{2}}$ is extrapolated
to be about $\unit[22]{T}$.

\section{Conclusions }

Our thermal-expansion, specific-heat, magnetostriction and magnetization
study allowed us to complete a comprehensive magnetic phase diagram
for $\mathrm{UAu_{2}Si_{2}}$. 

The magnetostriction curves measured at higher temperatures confirm
bulk character of the $\unit[50]{K}$ weak FM phase. The large volume
contraction in the UAFM ordered state suggests a large positive pressure
dependence of $T_{\mathrm{m}}$. The linear thermal-expansion data
point on the opposite effect for uniaxial pressure applied along the
tetragonal $c$ axis and within the basal plane. Magnetization measurements
in a hydrostatic pressure cell, however, revealed a negligible hydrostatic
pressure effect on $T_{\mathrm{m}}$, namely $\mathrm{d}T_{\mathrm{m}}/\mathrm{d}p\approx\unit[0.6(1)]{K\,GPa^{-1}}$
in pressures up to $\unit[1.0]{GPa}$. Small initial increase of $T_{\mathrm{m}}$
under hydrostatic pressure up to $\unit[1]{GPa}$ was observed on
the specific-heat data, while continuous decrease is found for higher
pressure up to $\unit[2.79]{GPa}$. These values are much smaller
than the prediction from the Ehrenfest relations ($\mathrm{d}T_{\mathrm{m}}/\mathrm{d}p\approx\unit[4.9(1)]{K\,GPa^{-1}}$).
Further complex studies involving hydrostatic and uniaxial pressure
would be desired to shed more light on the nature of this controversy. 

As the order of all the relative length changes is $\sim10^{-5}$,
we can expect some structural changes or distortions of the $\mathrm{UAu_{2}Si_{2}}$
unit cell in the ground state. Our comparative dilatometry measurements
of the linear thermal expansion along the $a$ axis and along the
$\left[110\right]$ direction clearly show the fourfold symmetry breaking
in the basal plane. This may also affect the real low-pressure dependence
of the ordering temperature. High-resolution diffraction measurements
are needed to find the ground-state space group. Possible candidates
are the orthorhombic non-isomorphic subgroups $Fmmm$ and $Immm$,
where the first one was found to describe the structure of the high-quality
samples of $\mathrm{URu_{2}Si_{2}}$ in the hidden-order state. Our
high-field magnetization measurements revealed a critical field of
$\approx\unit[22]{T}$ where the ordering temperature $T_{\mathrm{m}}$
is suppressed to $\unit[0]{K}$. The hysteresis of this transition
emerges at a tricritical point given by $T_{\mathrm{m}}\approx\unit[16]{K}$
and $\mu_{0}H_{\mathrm{m}}\approx\unit[16]{T}$ as a sign of the change
of the transition from second to first order. 
\begin{acknowledgments}
The present research was supported by JSPS KAKENHI Grant Numbers JP17K05525,
JP15KK0146, JP15K05882, JP15K21732, and the Strategic Young Researcher
Overseas Visits Program for Accelerating Brain Circulation from JSPS.
Single crystal growth and majority of experiments were performed in
the Materials Growth and Measurement Laboratory MGML (see: http://mgml.eu).
We acknowledge the support of HLD at HZDR, a member of the European
Magnetic Field Laboratory (EMFL), where the magnetization measurements
in high pulsed fields have been done. We would like to thank to Sergei
Zherlitsyn for his help during the high-field magnetization measurement. 
\end{acknowledgments}

\bibliographystyle{apsrev4-1}

\end{document}